\documentclass[3p,authoryear,a4paper]{elsarticle}
\usepackage{graphicx}
\usepackage{xcolor}
\usepackage{dsfont}
\usepackage{amsmath}
\usepackage{amssymb}
\usepackage{amsfonts}
\usepackage{amsbsy}
\usepackage{amsthm}
\usepackage{array}
\usepackage{booktabs}
\usepackage{verbatim}
\usepackage[english]{babel}
\usepackage{float}
\usepackage{epsfig}
\usepackage{multirow}
\usepackage{natbib}
\usepackage{hyperref}
\usepackage{float}
\usepackage[font=small,skip=3pt]{caption}
\usepackage{graphicx}
\usepackage{caption}
\usepackage{subcaption}
\usepackage{float}
\usepackage{wrapfig}
\usepackage{babel}
\usepackage{comment}
\usepackage{algorithm}
\usepackage{algpseudocode}
\usepackage{graphbox,graphicx}

\newcolumntype{C}[1]{>{\centering\let\newline\\\arraybackslash\hspace{0pt}}m{#1}}
\renewcommand\appendix{\par
\setcounter{section}{0}%
\setcounter{subsection}{0}%
\setcounter{table}{0}
\setcounter{figure}{0}
\gdef\thetable{\Alph{table}}
\gdef\thefigure{\Alph{figure}}
\gdef\thesection{\Alph{section}}
\setcounter{section}{0}}

\newcolumntype{L}[1]{>{\raggedright\let\newline\\\arraybackslash\hspace{0pt}}m{#1}}
\newcolumntype{C}[1]{>{\centering\let\newline\\\arraybackslash\hspace{0pt}}m{#1}}
\newcolumntype{R}[1]{>{\raggedleft\let\newline\\\arraybackslash\hspace{0pt}}m{#1}}

\newtheorem{remark}{Remark}[section]
\numberwithin{equation}{section}
\numberwithin{figure}{section}
\numberwithin{table}{section}

\newcommand{\revpast}[1]{\textcolor{black}{#1}}
\newcommand{\rev}[1]{\textcolor{black}{#1}}

\newenvironment{arcitem}{
\begin{list}{---}{
	\topsep=1pt 
	\itemsep=1pt 
	\parsep=0pt 
	\leftmargin=19pt 
}}
{\end{list}}

\newcounter{arclist}

\newcounter{arcenum}
\begin{document}

\begin{frontmatter}

\title{Ensemble distributional forecasting for insurance loss reserving}

\author[UM]{Benjamin Avanzi}
\ead{b.avanzi@unimelb.edu.au}

\author[UNSW]{Yanfeng Li \corref{cor}}
\ead{yanfeng.li@student.unsw.edu.au}

\cortext[cor]{Corresponding author. }

\author[UNSW]{Bernard Wong}
\ead{bernard.wong@unsw.edu.au}

\author[MQ]{Alan Xian}
\ead{alan.xian@mq.edu.au}

\address[UM]{Centre for Actuarial Studies, Department of Economics, University of Melbourne VIC 3010, Australia}

\address[UNSW]{School of Risk and Actuarial Studies, UNSW Business School, UNSW Sydney NSW 2052, Australia}

\address[MQ]{Department of Actuarial Studies and Business Analytics, Macquarie Business School, Macquarie University NSW 2109, Australia}

\begin{abstract}

Loss reserving generally focuses on identifying a single model that can generate superior predictive performance. However, different loss reserving models specialise in capturing different aspects of loss data. This is recognised in practice in the sense that results from different models are often considered, and sometimes combined. For instance, actuaries may take a weighted average of the prediction outcomes from various loss reserving models, often based on subjective assessments. 

In this paper, we propose a systematic framework to objectively combine (i.e. ensemble) multiple \textit{stochastic} loss reserving models such that the strengths offered by different models can be utilised effectively. 
Our framework contains two main innovations compared to existing literature and practice. Firstly, our criteria model combination considers the full distributional properties of the ensemble and not just the central estimate - which is of particular importance in the reserving context. Secondly, our framework is that it is tailored for the features inherent to reserving data. These include, for instance, accident, development, calendar, and claim maturity effects. Crucially, the relative importance and scarcity of data across accident periods renders the problem distinct from the traditional ensembling techniques in statistical learning. 

Our framework is illustrated with a complex synthetic dataset. In the results, the optimised ensemble outperforms both (i) traditional model selection strategies, and (ii) an equally weighted ensemble. In particular, the improvement occurs not only with central estimates but also relevant quantiles, such as the 75th percentile of reserves (typically of interest to both insurers and regulators). \rev{The framework developed in this paper can be implemented thanks to an R package, \texttt{ADLP}, which is available from CRAN.}

\end{abstract}


\end{frontmatter}

\section{Introduction} \label{intro}

\subsection{Background}

The prediction of outstanding claims is a crucial process for insurers, in order to establish sufficient reserves for future liabilities, and to meet regulatory requirements. Due to the stochastic nature of future claims payments, insurers not only need to provide an accurate central estimate of the reserve, but they also need to forecast the distribution of the reserve accurately. Furthermore, in many countries, provision of the predictive distribution of outstanding claims is required by regulatory authorities. For instance, the Australian Prudential Regulation Authority (APRA) requires general insurers to establish sufficient reserves to cover outstanding claims with at least $75\%$ probability \citep{gps340}. Therefore, insurers need to ensure that their models achieve satisfactory predictive performance beyond the central estimate (expectation) of outstanding claims. This is sometimes referred to as ``stochastic loss reserving'' \citep{WuMe08}.

Early stochastic loss reserving approaches include the distribution-free stochastic Chain-Ladder model proposed by \citet*{Ma93}. As computational power increased, Generalised Linear Models (GLMs) became the classical stochastic modelling approach \citep*{TaMG16}. They have been widely applied in both practice and loss reserving literature due to their stochastic nature and high flexibility in modelling various effects on claims payments, such as accident period effects, development period effects, calendar period effects, and effects of claims notification and closure. Common distribution assumption for GLMs include the log-normal distribution \citep*{Ver91}, over-dispersed Poisson (ODP) distribution \citep*{ReVe98}, and gamma distribution \citep*{Mac91}. Parametric curves to describe claim development patterns, such as the Hoerl curve and the Wright curve \citep*{Wri90} can also be incorporated in the GLM context. 

In recent years, machine learning models have become popular in loss reserving due to their potential in prediction accuracy as well as the potential for saving human labour costs. Notable examples include the smoothing spline \citep*{EnVe01}, Generalised Additive Model for Location, Scale and Shape (GAMLSS) \citep*{Sp14}, and Neural Networks \citep*{Wut18,Kuo19,AMAvTaWo22}.

Despite the sophisticated development of loss reserving models, the main focus of current loss reserving literature is to identify a \emph{single} model that can generate the best performance according to some measure of prediction quality. This approach is usually referred to as the ``model selection strategy''. However, due to the difference in model structures and assumptions, different loss reserving models are usually specialised at capturing specific claims effects, which are different from one model to the other. However, \emph{multiple} claims effects are typically present in reserving data. Furthermore, these effects are likely to appear inhomogenously accross accident and/or development years, which are noted in \cite*{Fri10} and \citet*[Chapter 5 of][pp. 151--165]{Tay00}. This makes it challenging for a single model to capture all of them. Using a single model to fit claims payments occurred in all accident periods may not be the most effective way to utilise the strengths offered by all the different available models.

Those concerns highlight the potential to develop a model \emph{combination} (as opposed to \emph{selection} strategy.

In practice, ensembles of different deterministic loss reserving models may have been typically constructed based on ad-hoc rules. The weights allocated to component models are commonly selected subjectively based on the model properties \citep*[see, e.g.,][]{Tay00,Fri10}. However, due to the subjective nature of this combination strategy, experience based on model properties may not generalise well to new claims data, and the weights allocated to component models are not optimised \citep*{Fri10}. Therefore, the information at hand may not be utilised most effectively. The subjective model combination strategy can further cause complications in cases where one wants information on the random distribution of claims. To address this, a more rigorous framework for combining stochastic loss reserving models was developed by \citet*[Chapter 12 of][pp. 151--165]{Tay00}, who proposed to select model weights by minimizing the variance of total estimated outstanding claims. While a great step forward, this approach only considers the first and the second moments of the claims distribution. This may not be sufficient to derive the required quantiles from the distribution of reserves accurately. 

Beyond the loss reserving literature, stacked regression has been recently applied in combining mortality forecasts \citep*{KeShViZi22, Li22}. Under the stacked regression approach, the combination weights are determined by minimising the Mean Squared Error (MSE) achieved by the ensemble. Despite showing success for combining point forecasts, this approach does not take into account the ensemble's performance over the entire distribution, which is critical for loss reserving purposes. 

The so-called ``linear pool'' \citep*{GnRaWeGo05}, which has been widely applied in combining distributional forecasts outside the field of loss reserving, may offer an alternative model combination approach that could overcome the aforementioned limitations of model selection strategy, and of traditional combination strategies for loss reserving models. The linear pool uses a \emph{proper scoring rule} to determine the weights allocated to component distributional forecasting models. A proper scoring rule assigns a numeric score to a distributional forecasting model when an event materialises. This presents a number of advantages when assessing the quality of distributional forecasts \citep{GnRa07}. Literature in the areas of weather forecast \citep{GnRaWeGo05, RaGnBaPo05, GnRa07, BaMa05, VoKnFiScGn18}, stocks return forecast, and GDP growth rate forecast \citep{HaMi07,McTh11,OpVaVan17,HaMi04} all suggest that linear pool ensembles generate more accurate distributional forecast than any single model in their respective contexts. A common choice for the proper score is the Log Score. Furthermore, linear pools are akin to finite mixture models in producing a weighted average of predictive distributions, yet they differ in weight determination. Finite mixture models simultaneously train model parameters and weights using the same dataset \citep*{BaLe18}. Conversely, the linear pooling method first calibrates model parameters on a training set and then optimises combination weights on a separate validation set. This is likely to improve the ensemble's overall predictive accuracy compared to standard finite mixture models.
In this paper, we start by developing a linear pool approach to loss reserving, and extend the methodology to best fit the reserving context, as detailed in the following section.

\subsection{Statement of contributions} \label{StatementContri}

To our knowledge, linear pools have not appeared in the loss reserving literature. Unfortunately, existing linear pool implementations cannot be applied in this context directly for a number of reasons outlined below. This paper aims to fill this gap by developing an adjusted linear pool ensemble forecasting approach that is particularly suitable to the loss reserving context. Although actuaries may combine different loss reserving models in practice, we argued earlier that the model combination is then  generally performed subjectively and may not optimise the ensemble's performance. Compared with the subjective model combination approach, our proposed scheme has two advantages. Firstly, data driven weights reduce the amount of manual adjustment required by the actuary, particularly in cases where a new experience has emerged. Secondly, by using a proper score as the objective function, the distributional forecast accuracy of the ensemble can be optimised. As \citet*{Sh22} briefly suggests in his monograph, the future development of model combination strategies for loss reserving models should consider the relative predictive power of component models, and the weighted results should ``reflect the actuary’s judgements about the entire distribution, not just a central estimate''. Since the proper score can formally assess the quality of distributional forecast, the linear pools satisfy the two conditions proposed by \citep*{Sh22} by using a proper score as the optimisation criterion.

In this paper, we introduce a distributional forecast combination framework that is \emph{tailored} to the loss reserving context. Although the standard ensemble modelling approach has been extensively studied in the literature, it can not be applied directly without modification due to the special data structures and characteristics of the loss reserving data. Specifically, we introduce ``\textbf{Accident and Development period adjusted Linear Pools (ADLP)}'', which have the following attributes:

\begin{arcitem}

    \item \textit{Triangular shape of aggregate loss reserving data}: To project out-of-sample incremental claims, the \textbf{training data must have at least one observation from each accident period and development period} in the upper triangle. Our proposed ensemble ensures that this constraint is satisfied when partitioning the training and validation sets. 
    
    \item \textit{Time-series characteristics of incremental claims}: Under the standard ensemble approach, the in-sample data is \emph{randomly} split into the training set for training component models and the validation set for estimating combination weights \citep*{Br96}. 
    
    However, since the main purpose of loss reserving is to predict future incremental claims, a random split of the data can not adequately validate the extrapolating capability of loss reserving models. In light of the above concern, we \textbf{allocate incremental claims from the most recent calendar periods to the validation set} (rather than a randomly selected sample) as they are more relevant to the future incremental claims. This is a common partition strategy used in loss reserving literature \citep*[e.g.,][]{TaMG16}. Therefore, the combination weights estimated from the validation set more accurately reflect the component models' predictive performance on the most recent out-of-sample data. 
    
    \item \textit{Accident Period effects}: The standard ensemble strategy usually derives its combination weights by using the whole validation set, which implies each component model will receive a single weight for a given data set. 
    
    However, it is a very well known fact that the performance of loss reserving models tends to differ by accident periods \citep*[Chapter 5 of][pp. 151--165]{Tay00}. Hence, assigning the same weights (for each model) across \emph{all} accident years might not be optimal, and it may be worthwhile allowing those to change over time. To address this, we allow the set of weights to change at arbitrary ``split points''. For instance, the set might be calibrated differently for the 10 most recent (immature) accident years. 
    
    More specifically, we further \textbf{partition the validation set based on accident periods to train potentially different combination weights}, such that the weights can best reflect a models' performance in the different accident period sets. Our paper shows that the ensemble's  out-of-sample performance can be substantially improved---based on the whole range of evaluation metrics considered here---by adopting the proposed Accident-Period-dependent weighting scheme. Additionally, we also empirically discuss the choice of number and location of accident year split points . 
    
    \item  \textit{Development Period effects}: 
    
    The accident year split described in the previous item requires some care. One cannot strictly isolate and focus on a given set of accident years (think of a horizontal band) and use the validation data of that band only as this will lack data from later development years which are required for the projection of full reserves. Hence, we start with the oldest accident year band (or split set), and when moving to the next band, we (i) change the training set according to the accident year effect argument discussed above, but (ii) add the new layer of validation points to the previously used set of validation data when selecting weights for that band.
    
   Note that the data kept from one split to the next is the validation data only, which by virtue of the second item above is always the most recent (last calendar year payments). Hence, this tweak does not contradict the probabilistic  forecasting literature's recommendation of using only the data in the most recent periods \citep*[see, e.g.][]{OpVaVan17} when selecting weights.
    
   \item \textit{Data Scarcity}: Data is scarce for aggregate loss reserving, particularly for immature accident periods (i.e., the recent accident periods), where there are only a few observed incremental claims available. A large number of component models relative to the number of observations may cause the prediction or weights estimation less statistically reliable \citep*{AiTi06}. Our proposed ensemble scheme \textbf{mitigates this data scarcity issue by incorporating incremental claims from both mature and immature accident periods to train the model weights in immature accident periods}.

 \end{arcitem}

The out-of-sample performance of the ADLP is evaluated and compared with single models (model selection strategy), as well as other common model combination methods, such as the equally weighted ensemble, by using synthetic data sets that possess complicated features inspired by real world data.  

Overall, we show that our reserving-oriented ensemble modelling framework ADLP yields more accurate distributional forecasts than either (i) a model selection strategy, or (ii) the equally weighted ensemble. The out-performance of the proposed ensembling strategy can be credited to its ability to effectively combine the strengths offered by different loss reserving models and capture the key characteristics of loss reserving data as discussed above. In fact, the mere nature of the challenges described above justify and explain why an ensembling approach is favourable. \rev{Finally, we have also developed an R package \texttt{ADLP}, which has been published in CRAN, to promote the usability of the proposed framework.}

\subsection{Outline of the paper}

In Section \ref{Component}, we discuss the design and properties of component loss reserving models that will constitute the different building blocks for the model combination approaches discussed in this paper. Section \ref{modCombine} discusses how to best combine those component models, and describes the modelling frameworks of the standard linear pool ensemble and the Accident and Development period adjusted Linear Pools (ADLP) that are tailored to the general insurance loss reserving data. The implementation of our framework is not straighforward, and is detailed in Section \ref{S_implem}. 
Section \ref{comparepred} discusses the evaluation metrics and statistical tests used to assess and compare the performance of different models. The empirical results are illustrated and discussed in Section \ref{empiricalResults}. Section \ref{conclusions} concludes.  

\section{Choice of component models} \label{Component}

Given the extensive literature in loss reserving, which offers a plethora of modelling options, we start by defining criteria for choosing models to include in the ensemble---the ``component models''. We then describe the structure of each of the component models we selected, as well as the characteristics of loss data that they aim to capture.

\subsection{Notation} \label{Notations}

We first define the basic notation that will used in this paper:
\begin{itemize}
    \item $i$: the index for accident periods, where $i = 1,...,I$
    \item $j$: the index for development periods, where $j = 1,...,J$ and $J=I$
    \item $t$: the calendar period of incremental claims; $t=i+j-1$, where $i$ and $j$ denote the accident period and the development period, respectively
    \item $D_{out}$: the out-of-sample loss data (i.e. the lower part of an aggregate loss triangle)
    \item $D_{Train}$: the loss data used for training component models
    \item $D_{val}$: the validation loss data for selecting component models and optimizing the combination weights; $D_{Train}$ and $D_{val}$ constitute the in-sample loss data (i.e. the upper part of an aggregate loss triangle)
    \item $Y_{ij}$: the observed incremental claims paid in accident period $i$ and development period $j$
    \item $\hat{Y}_{ij}$: the estimated incremental claims in accident period $i$ and development period $j$
    \item $R^{\text{True}}$: the true reserve, defined as $R^{\text{True}}=\sum_{i,j \in D_{out}} Y_{ij}$
    \item $\hat{R}$: the estimated reserve, defined as $\hat{R}=\sum_{i,j \in D_{out}} \hat{Y}_{ij}$
    \item $N_{ij}$: the observed reported claims count occurred in accident period $i$ and development period $j$
    \item $F_{ij}$: the observed number of finalised claims in accident period $i$ and development period $j$
    \item $\hat{\mu}_{ij}$: the predicted mean for the distribution of $Y_{ij}$ (or $\ln(Y_{ij})$ for the Log-Normal distribution)
    \item $\hat{\sigma}_{ij}^2$: the predicted variance for the distribution of $Y_{ij}$ (or $\ln(Y_{ij})$ for the Log-Normal distribution)
    \item $\phi$: the dispersion parameter for the distribution of $Y_{ij}$; for distributions coming from Exponential Distribution Family (EDF), $\text{Var}(Y_{ij})$ is a function of $\phi$ and $\mu_{ij}$
    \item $\eta_{i,j}$: the linear predictor for incremental claims paid in accident period $i$ and development period $j$ that incorporates information from explanatory variables into the model; $\eta_{i,j}=h^{-1}(\hat{\mu}_{ij})$, where $h^{-1}$ is a link function 
    \item $\hat{f}_m(y_{ij})$: the predicted density by model $m$ evaluated at the observation $Y_{ij}$
    \item $m$: the index for component models, where $m=1,...,M$
\end{itemize}

\subsection{Criteria for choosing component models} \label{crit}

There is a wide range of loss reserving models available in the literature. With an aim to achieve a reasonable balance between diversity and computational cost, we propose the following criteria to select which component models to include in our ensembles: 

\begin{enumerate}
    \item 
    \emph{The component models should be able to fit mechanically without requiring substantial user adjustment in order to avoid the potential subjective bias in model fitting, and to save time spent on manually adjusting each component model.}
    
    This can be of particular concern when there is a large number of component models contained in the ensemble. This criterion is consistent with the current literature on linear pool ensembles, where parametric time-series models with minimum manual adjustments being required are used as component models \citep*{GnRa13,OpVaVan17}.
    \item 
    \emph{The component models should have different strengths and limitations so as to complement each other in the ensemble.} 
    
    This criterion ensures that different patterns present in data can be potentially captured by the diverse range of component models in the ensemble. As \citet*{Br96} suggests, having diverse component models could also reduce the potential model correlation, thus improving the reliability and accuracy of the prediction results. 
    \item 
    \emph{The component models should be easily interpretable and identifiable, and hence are restricted to traditional stochastic loss reserving models and statistical learning models with relatively simple structures.} 
    
    Although advanced machine learning models, such as Neural Networks, have demonstrated outstanding potential in forecast accuracy in literature \citep*{Kuo19}, they might place a substantial challenge on the interpretability of the ensemble and the data size requirement if they are included in the ensemble. Therefore, their implementation into the proposed ensemble has not been pursued in this paper.
\end{enumerate}

In summary, the ensemble should be automatically generated (1), rich (2) and constituted of known and identifiable components (3).

Note, however, that our framework works for any number and type of models in principle, provided the optimisation described in Section \ref{modCombine} can be achieved.

\subsection{Summary of selected component models}

Based on the criteria for choosing component models introduced in the previous section,
we have selected eighteen reserving models with varying distribution assumptions and model structures from the literature. These include GLM models with various effects and various dispersions, zero-adjusted models, smoothing splines, as well as GAMLSS models with varying dispersions.  While Appendix \ref{Appendix:ComponentModels} provides a detailed description of those models, Table \ref{table:SumComponentMod} summarises the common patterns in claim payments data that actuaries could encounter in modelling outstanding claims, and the corresponding component models that can capture those patterns.

\begin{table}[htb]
\footnotesize
\centering

\begin{tabular}{ccccc}

\textbf{Model Class}    & \textbf{Model Structures} & \textbf{Fitted Distributions} & \textbf{Effects}  \\ \hline
    \multirow{5}{*}{GLM Based Model} &$\text{GLM}_{\text{CC}}$ & ODP, Log-Normal, Gamma & accident and development effects \\ &$\text{GLM}_{\text{Cal}}$ & ODP, Log-Normal, Gamma & calendar period effects\\ & $\text{GLM}_{\text{HC}}$ & ODP, Log-Normal, Gamma &shape of development patterns\\ & $\text{GLM}_{\text{PPCI}}$ &ODP &claims notification effects \\ & $\text{GLM}_{\text{PPCF}}$ & ODP & claims finalisation effects\\
    & $\text{GLM}_{\text{ZALN}}$, $\text{GLM}_{\text{ZAGA}}$ &ZALN, ZAGA & occurrence of zero incremental claims \\
    \hline 
    Smoothing Splines &$\text{SP}$ & Normal, Log-Normal, Gamma & shape of development patterns \\
    \hline 
    GAMLSS &$\text{GAMLSS}$ & Log-Normal, Gamma & varying dispersion \\

    \hline
\end{tabular}

\label{tab:multicol}
\caption{Summary of component models\label{table:SumComponentMod}}
\end{table}

\begin{remark}
    Another group of models that are easily interpretable and could be fitted mechanically is the age-period-cohort type of models proposed in \cite*{HaNi18} and \cite*{KuNiNi08}. \revpast{These models have a close connection to the model with calendar period effects (i.e., $\text{GLM}_{\text{Cal}}$) and the cross--classified model (i.e., $\text{GLM}_{\text{CC}}$), which will be discussed in more details in Appendix \ref{CalPeriod}.}
   
\end{remark}

\section{Model combination strategies for stochastic ensemble loss reserving} \label{modCombine}

In this section, we first define which criterion we will use for determining the relative strength of one model when compared to another. We then review the two most basic ways of using models within an ensemble: (i) the Best Model in the Validation set (BMV), and (ii) the Equally Weighted ensemble (EW). Both are deficient, in that (i) considers the optimal model, but in doing so discards all other models, and (ii) considers all models, but ignores their relative strength. To improve these and combine the benefits of both approaches, both the Standard Linear Pool (SLP) and Accident and Development period adjusted Linear Pools (ADLP) are next developed and tailored to the reserving context. The section concludes with details of the optimisation routine we used, as well as how predictions were obtained.

\subsection{Model combination criterion} \label{modEval}

\subsubsection{Strictly proper scoring rules and probabilistic forecasting}

We consider ``\emph{strictly proper}'' scoring rules, which have wide application in probabilistic forecasting literature \citep*{GnKa14}, to measure the accuracy of distributional forecast. Scoring rules assign an numeric score to a distributional forecasting model when an event materialises. Define $S(P,x)$ as the score received by the forecasting model $P$ when the event $x$ is observed, and $S(Q,x)$ as the score assigned to the true model $Q$. A scoring rule is defined to be ``\emph{strictly proper}'' if $S(Q,x) \ge S(P,x)$ always holds, and if the equality holds if and only if $P=Q$. The definition of proper scoring rules ensures the true distribution is always the optimal choice, thus encouraging the forecasters to use the true distribution \citep*{GnRa07,GnKa14}. This property is crucial as improper scoring rules can give misguided information about the predictive performance of forecasting models \citep*{GnRa07,GnKa14}.

Furthermore the performance of a distributional forecasting model can be evaluated by its calibration and sharpness. Calibration measures the statistical consistency between the predicted distribution and the observed distribution, while sharpness concerns the concentration of the predicted distribution. A general goal of distributional forecasting is to maximise the sharpness of the predictive distributions, subject to calibration. By using a proper scoring rule, the calibration and sharpness of the predictive distribution can be measured at the same time. For introduction to scoring rules and the advantages of using strictly proper scoring rules, we refer to \citep*{GnRa07,GnKa14}. 

Actuaries commonly employ goodness-of-fit criteria such as the Kolmogorov-Smirnov or Anderson-Darling statistics, to assess the level of agreement between theoretical distributions and empirical distributions. However, it is crucial to recognise the fundamental distinction between evaluating probabilistic forecasts and assessing the goodness-of-fit of a theoretical distribution. In probabilistic forecasting, our objective is to evaluate the predictive distribution based on a single value, as the true distribution of a particular instance is typically unobservable in practical scenarios. Given this consideration, proper scoring rules are better suited for our purposes.

\subsubsection{Log Score}

In this paper, we use the Log Score, which is a strictly proper scoring rule proposed by \citet*{Go92}, to assess our models' distributional forecast accuracy. Log score is also a local scoring rule, which is any scoring rule that depends on a density function only through its value at a materialised event \citep*{MaPhSt12}. Under certain regularity conditions, it can be shown that every proper local scoring rule is equivalent to the Log Score \citep*{GnRa07}. For full discussion of the advantages of Log Score, we refer to \citep*{RoSm02, GnRa07}.

For a given dataset D, the average Log Score attained by a predictive distribution can be expressed as:
 \begin{eqnarray}
\text{LogS} &=& \frac{1}{|D|}\sum_{i,j \in D}\ln(\widehat{f}(y_{i,j})),
\end{eqnarray}
where $\widehat{f}(y_{i,j})$ is the predicted density at the observation $y_{ij}$, and $|D|$ is the number of observations in the dataset D.

\begin{remark}
Here, the Log Score is chosen to calibrate model component weights in the validation sets (the latest diagonals). Strictly proper scoring rules are also used to assess the quality if the distributional forecasting properties of the ensembles developed in this paper, but this time applied in the test sets (in the lower triangle). This is further described in Section \ref{comparepred}.
\end{remark}

\subsection{Best Model in the Validation set (BMV)}

Traditionally, a single model, usually the one with the best performance in the validation set, is chosen to generate out-of-sample predictions. In this paper, the Log Score is used as the selection criterion, and the selected model is referred to as the ``\emph{best model in validation set}''---abbreviated ``\textbf{BMV}''. The model selection strategy can be regarded as a special ensemble where a weight of $100\%$ is allocated to a single model (i.e. the \textbf{BMV}). Although model selection is based on the relative predictive performance of various models,  it is usually hard for a single model to capture all the complex patterns that could be present in loss data \citep*{Fri10}. In our alternative ensembles, we aim to incorporate the strengths of different models by combining the prediction results from various models.

\subsection{Equally Weighted ensemble (EW)}

The simplest \emph{model combination} strategy is to assign equal weights to all component models, which is commonly used as a benchmark strategy in literature \citep*{RaGn10, HaMi07, McTh11}. For this so-called ``\emph{equally weighted ensemble}''---abbreviated ``\textbf{EW}'', the combined predictive density can be specified as
\begin{eqnarray}
f_{*}(y_{ij}) &=& \sum_{m=1}^{M}\frac{1}{M}\widehat{f}_m(y_{ij}),
\end{eqnarray}
where $M$ is the total number of component models, and $\widehat{f}_m(y_{ij})$ is the predicted density from the component model $m$ evaluated at the observation $y_{ij}$. As the equally weighted ensemble can sometimes outperforms more complex model combination schemes \citep*{ClMaVaWa16, SmWa09, PiStYa12}, we also consider it as a benchmark model combination strategy in this paper. 

\subsection{Standard linear pool ensembles (SLP)} \label{impleLP}

A more sophisticated, while still intuitive approach is to let the combination weights be driven by the data so that the ensemble's performance can be optimised. A prominent example of such combination strategies in probabilistic forecasting literature is the linear pool approach \citep*{GnRa13}, which seeks to aggregate individual predictive distributions through a linear combination formula. The combined density function of such ``\emph{standard linear pool}''---abbreviated ``\textbf{SLP}'' can be then specified as
\begin{eqnarray} \label{ensPredict}
f^{*}(y_{ij}) &=& \sum_{m=1}^{M}w_m \cdot f_m(y_{ij})
\end{eqnarray}
at observation $y_{ij}$, where $w_m$ is the weight allocated to predictive distribution $m$ \citep*{GnRa13}. The output of linear pools, as defined in \eqref{ensPredict}, closely resembles that of finite mixture models, with both being a weighted average of an ensemble of predictive distributions. However, unlike finite mixture models, the calibration of a linear pool ensemble usually involves two stages. In the first stage, the component models are fitted in the training data, and the fitted models will generate predictions for the validation set, which constitute the Level 1 predictions. In the second stage, the optimal combination weights are learned from the Level 1 predictions by maximizing (or minimising) a proper score. Since the weights allocated to component models should reflect their out-of-sample predictive performance, it is important to determine the combination weights by using models' predictions for the validation set instead of the training set. 

\begin{figure}[htb]
    \centering
    \includegraphics[width=\textwidth]{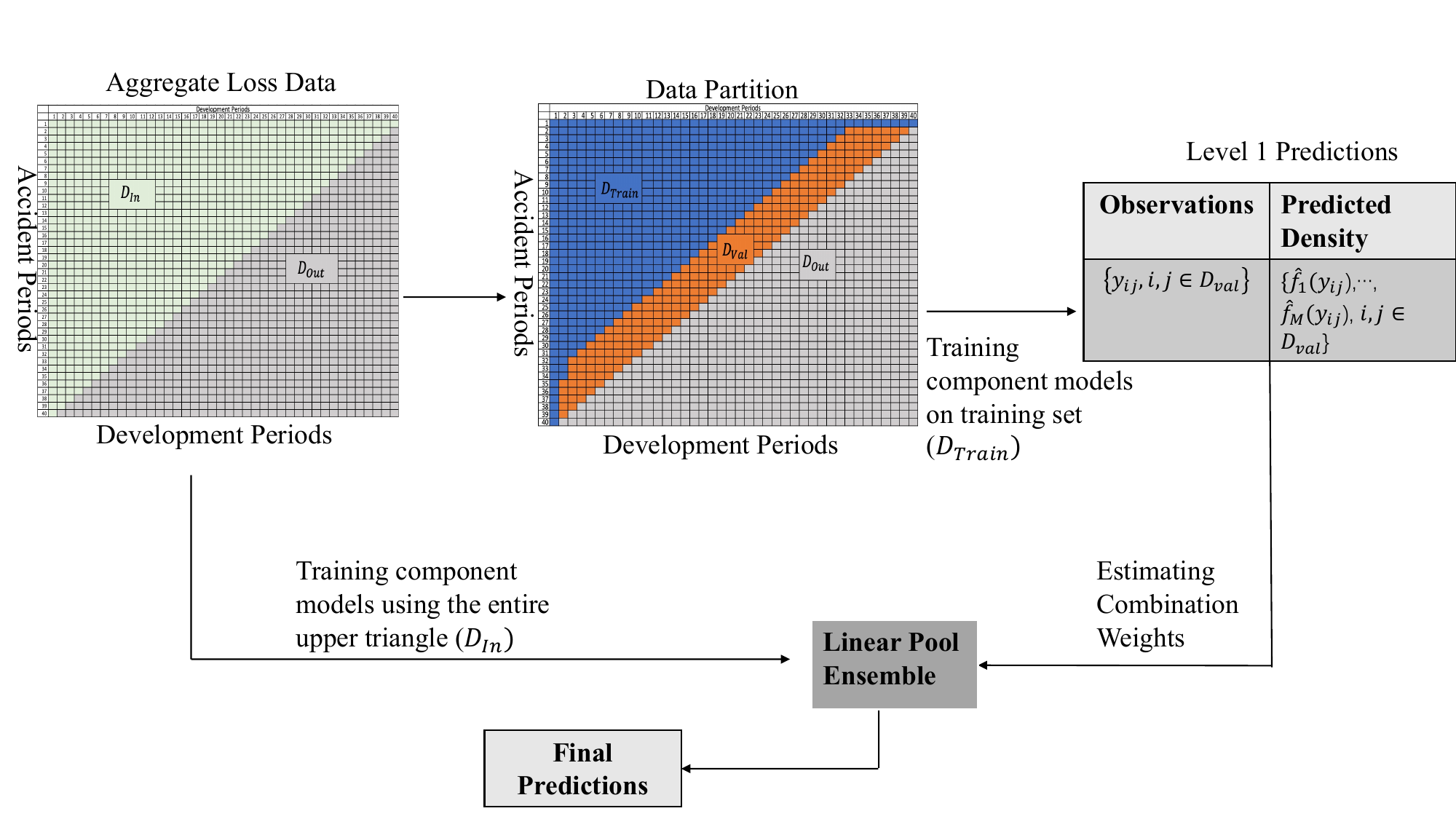}
    \caption{Linear Pool Ensemble Framework}
    \label{fig:IllustrationofLinearPool}
\end{figure}

Figure \ref{fig:IllustrationofLinearPool} illustrates the modelling framework for constructing a standard linear pool (SLP) ensemble, with the following steps: 
\begin{enumerate}
    \item Partition the in-sample data (i.e. the upper triangle $D_{in}$) into training set $D_{Train}$ (shaded in blue) and validation set $D_{val}$ (shaded in orange). We allocate the latest calendar periods to the validation set to better validate the projection accuracy of models, which is also  common practice in the actuarial literature.
    \item Fit the $M$ component models to the training data (i.e. $D_{Train}$), that is, find the model parameters that maximise the Log Score within $D_{Train}$ (equivalent to Maximum Likelihood estimation).
    \item Use the fitted models to generate predicted densities for incremental claims in the validation set, which forms the Level 1 predictions; Denote the Level 1 predictions as $\{\hat{f}_1(y_{ij}),...,\hat{f}_M(y_{ij}),i,j\in D^{val}\}$
    \item Estimate the optimal combination weights from the Level 1 predictions by maximizing the mean Log-Score achieved by the ensemble, subject to the constraint that combination weights must sum up to one (i.e. 
    $\sum_{m=1}^{M} w_m=1$) and each weight should be non-negative (i.e. $w_m \ge 0$). We have then
    \begin{eqnarray} \label{EqnOptimalProblem}
\hat{\textbf{w}} &=& \underset{\textbf{w}}{\mathrm{arg\,max}}\; \frac{1}{|D^{val}|}\sum_{y_{ij} \in D^{val}} \ln\left(\sum_{m=1}^{M} w_{m} \hat{f}_m(y_{i,j})\right),
\end{eqnarray}
where $\hat{\textbf{w}}=[\hat{w}_1,...,\hat{w}_M]'$. 

The sum-to-unity and non-negativity constraint on model weights are important for two main reasons. Firstly, those constraints ensure the ensemble not to perform worse than the worst component model in out-of-sample data (proof in Appendix \ref{proofConstraints}). Additionally, the non-negativity constraint can mitigate the potential issue of model correlation by shrinking the weights of some of the highly correlated models towards zero \citep*{CoGui20,Br96}. 
    \item Fit the component models in the entire upper triangle $D_{in}$, and then use the fitted models to generate predictions for out-of-sample data $D_{out}$. The predictions are combined using the optimal combination weights determined in step 4. 
\end{enumerate}

\subsection{Accident and Development period adjusted Linear Pools (ADLP) driven by general insurance characteristics} \label{LogScoreOW}

The estimation of the SLP combination weights specified in \eqref{EqnOptimalProblem} implies that each model will receive the same weight in all accident periods. However, as previously argued, loss reserving models tend to have different predictive performances in different accident periods, making it potentially ideal for combination weights to vary by accident periods. In industry practice, it is common to divide the aggregate claims triangle by claims maturity and fit different loss reserving models accordingly \citep*{Fri10}. 

For instance, the PPCF model tends to yield better predictive performance in earlier accident periods (i.e., the more mature accident periods) by taking into account the small number of outstanding claims. 
In contrast, the PPCI model and the Chain-Ladder tend to have better performance in less mature accident periods as they are insensitive to claims closure \citep*[Chapter 5 of][pp. 151--165]{Tay00}. Therefore, in practice, actuaries may want to assign a higher weight to the PPCF model in earlier accident periods and larger weights to the PPCI or Chain-Ladder model in more recent accident periods \citep*[Chapter 5 of][pp. 151--165]{Tay00}.

The argument above can be applied to other loss reserving models. To take into account  the accident-period-dependent characteristics of loss reserving models, we partition the loss reserving data by claims maturity, with the first subset containing incremental claims from earlier accident periods (i.e., the mature data) and the second subset containing incremental claims from more recent accident periods (i.e., the immature data). The data is thus split an arbitrary number of times $K$. The weight allocated to component model $m$ for projecting future incremental claims in the $k
^{th}$ subset (i.e. $D_{test}^{k}$) is then optimised using the $k^{th}$ validation subset (denoted as $D_{val}^{k}$): 
\begin{eqnarray} \label{EqnOptimalSubOne}
\hat{\textbf{w}}^{k} &=& \underset{\textbf{w}^k}{\mathrm{arg\,max}}\; \frac{1}{|D_{val}^{k}|}\sum_{y_{ij} \in D_{val}^{k}} \ln\left(\sum_{m=1}^{M} w_m^k \hat{f}_m(y_{i,j})\right),\quad, k=1,\ldots,K,
\end{eqnarray}
subject to $\sum_{m=1}^{M} w_m^k=1, w_m^k \ge 0$. The choice of split points, as well as $D_{val}^{k}$ require care, as discussed below. We will mainly focus on the case $K=2$, but this can be easily extended to $K>2$; see Remark \ref{Threesubsets} below and Appendix \ref{ADLPThreeSubsets}, which consider $K=3$ and briefly discuss how $K$ could be chosen.

\begin{table}[htb]
\centering \footnotesize
\caption{Details of Two-Subsets Data Partition Strategies. \label{OldParDetails}}
\begin{tabular}{llll}
Partition Strategies    & Accident Periods in Subset 1  & Accident Periods in Subset 2   & Proportion in Subset 1    \\
$1$                       & 2-3                          & 4-40                          & 14\%    \\
$2$                       & 2-4                          & 5-40                          & 20\%    \\
$3$                       & 2-5                          & 6-40                          & 23\%    \\
$4$                       & 2-7                          & 8-40                          & 32\%    \\
$5$                       & 2-9                          & 10-40                          & 40\%    \\
$6$                      & 2-11                          & 12-40                          & 47\%    \\
$7$                       & 2-13                          & 14-40                          & 54\%    \\
$8$                       & 2-14                          & 15-40                          & 57\%    \\
$9$                       & 2-15                          & 16-40                          & 60\%   \\
$10$                       & 2-16                          & 17-40                          & 63\%    \\
$11$                       & 2-17                          & 18-40                          & 66\%   \\ 
$12$                       & 2-18                          & 19-40                          & 69\%   \\
$13$                       & 2-19                         & 20-40                          & 72\%   \\
$14$                       & 2-23                          & 24-40                          & 81\%   \\
$15$                       & 2-26                          & 27-40  & 87\%   \\
$16$                       & 2-28                          & 29-40  & 90\%   \\
$17$                       & 2-31                         & 32-40  & 95\%   \\
$18$                       & 2-33                          & 34-40  & 97\%   \\
\end{tabular}
\end{table}

Table \ref{OldParDetails} summarises multiple ensembles constructed by using different split points between the two subsets and the corresponding proportion of data points in the upper triangle in the subset 1. We do not test split points beyond accident period 33 due to the scarcity of data beyond this point. 
Results for all those options will be  compared in Section \ref{empiricalResults}, with the objective of choosing the optimal split point.

\begin{remark}
    The ``empiricalness'' of the choice of split point might be seen as a limitation. It is hard to avoid as there might not be a closed formula for the best split point. While this does require extra work each period, the additional computation time is not significant. Also, many reserving models are iterative, suggesting minimal shifts in optimal partitions with a period’s worth of extra experience. This would suggest that a relatively stable choice of split point can be made early in the analysis, and maintained over future periods, assuming consistent historical experience.
\end{remark}

\begin{figure}[htb]
\centering
\includegraphics[width=0.6\textwidth]{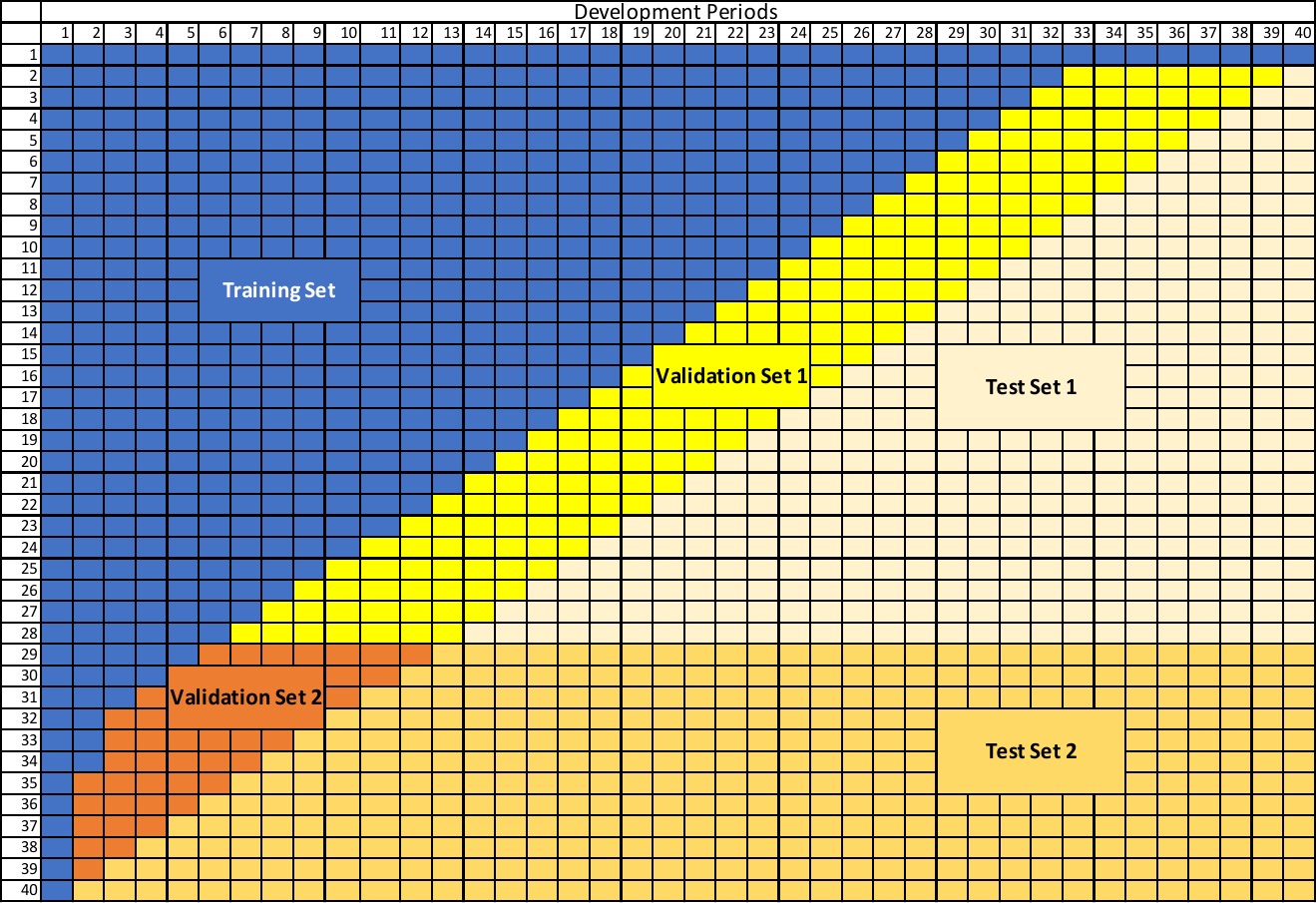}
\caption{Data Partition Diagram: An illustrative example for ensembles that only consider impacts from accident periods \label{ParDiagOne}}
\end{figure}

We now turn our attention to the definition of the validation sets to be used in \eqref{EqnOptimalSubOne}. This pure ``band'' approach leads to the definitions depicted in Figure \ref{ParDiagOne}. 
Unfortunately this may ignore very important development period effects, because the second validation set (in orange) does not include any of the development periods beyond 12. 

More specifically, the training scheme for combination weights specified by \eqref{EqnOptimalSubOne} implies the weights in Subset 2 (i.e. $\hat{\textbf{w}}^{2}$) only derive from Subset 2 validation data (i.e. $D_{val}^{2}$). However, 
$D_{val}^{2}$ does not capture the late development period effects in out-of-sample data: as shown in Figure \ref{ParDiagOne}, $D_{val}^{2}$ (shaded in orange) contains incremental claims from Development Period 2 to 12. However, $D_{test}^{2}$ (shaded in light orange) comprises incremental claims from Development Period 2 to 40. Depending on the lines of business, the impact from late development periods on claims could be quite significant. For example, for long-tailed lines of business, such as the Bodily Injury liability claims, it usually takes a considerably long time for the claims to be fully settled, and large payments can sometimes occur close to the finalisation of claims \citep*[Chapter 4 of][pp. 151--165]{Tay00}. In such circumstances, adequately capturing the late development period effects is critically important as missing those effects tends to substantially increase the risk of not having sufficient reserve to cover for those large payments that could occur in the late development periods. 

This situation can be exacerbated by the data scarcity issue in $D_{val}^2$ for ensembles with late split points. For instance, partition strategy 17 only uses claims from Accident Period 34 to 40 in $D_{val}^2$, which corresponds to only 20 observations. However, those 20 observations are used to train the 18 model weights. When the data size is small relative to the number of model weights that need to be trained, the estimation of combination weights tend to be less statistically reliable due to the relatively large sampling error \citep*{AiTi06,Be66}. 

In light of the above concerns, we modify the training scheme such that the mature claims data in $D_{val}^{1}$, which contains valuable information about late development period effects, can be utilised to train the weights in less mature accident periods. Additionally, since both $D_{val}^{1}$ and $D_{val}^{2}$ are used to train the weights in the second subset, the data scarcity issue in $D_{val}^{2}$ can be mitigated. Under the modified training scheme, \eqref{EqnOptimalSubOne} is revised as
\begin{eqnarray} \label{EqnOptimalSubTwo}
\hat{\textbf{w}}^{k} &=& \underset{\textbf{w}^k}{\mathrm{arg\,max}}\;
\frac{1}{|D_{val}^{1} \cup...\cup D_{val}^{k}|}\sum_{y_{ij} \in D_{val}^{1} \cup...\cup D_{val}^{k}} \ln\left(\sum_{m=1}^{M} w_m^k \hat{f}_m(y_{i,j})\right),\quad k=1,\ldots,K,
\end{eqnarray}
subject to $\sum_{m=1}^{M} w_m^k=1, w_m^k \ge 0$. Since the combination weights estimated using \eqref{EqnOptimalSubTwo} take account into impacts and features related to different Accident and Development period combinations, we denote the resulting ensembles as the \textbf{Accident and Development period adjusted Linear Pools (ADLP)}. 

\begin{figure}[htb]
\begin{minipage}{\textwidth}
\begin{minipage}[b]{0.5\textwidth}
    \centering
    \includegraphics[width=\textwidth]{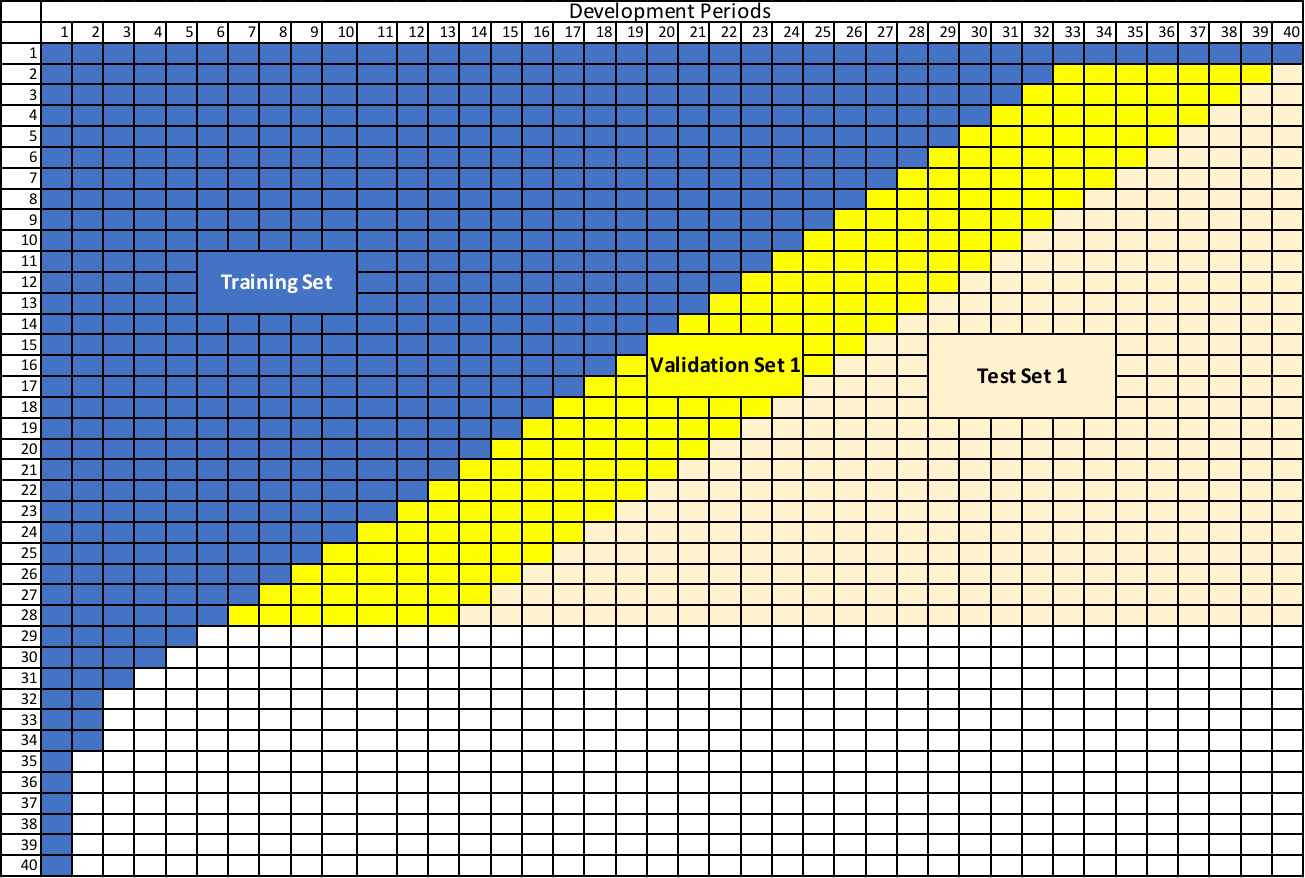}
    \captionof{figure}{ADLP Validation Subset 1}
    \label{ADLPValidSet1}
    \end{minipage}
  \begin{minipage}[b]{0.5\textwidth}
    \centering
    \includegraphics[width=\textwidth]{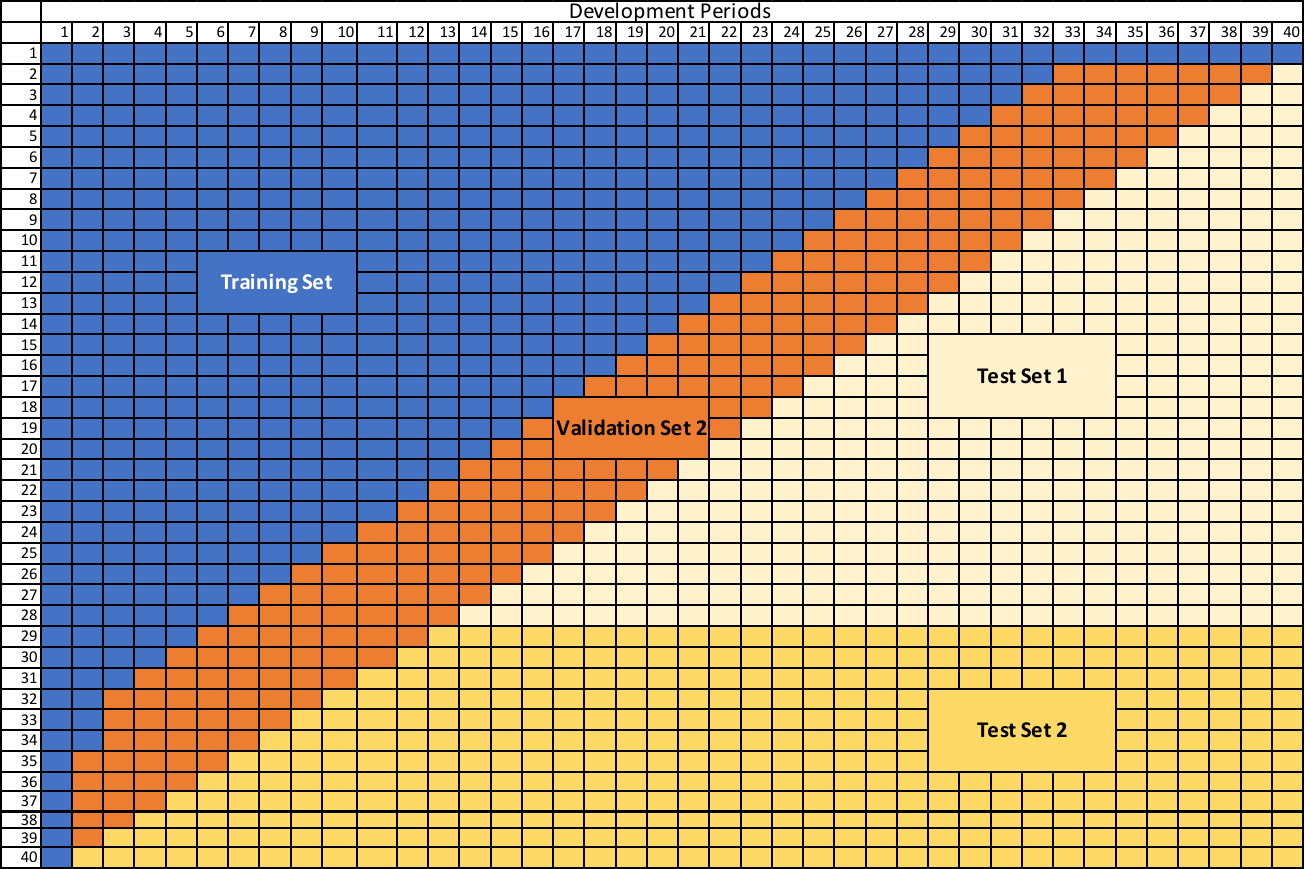}
    \captionof{figure}{ADLP Validation Subset 2}
    \label{ADLPValidSet2} 
  \end{minipage}
  \hfill
  \end{minipage}
\end{figure}

An example of data partition strategy under ADLP ensembles with $K=2$ is given in Figure \ref{ADLPValidSet1} and \ref{ADLPValidSet2}, where the new second validation set overlaps the previous first and the second validation set. Since the bottom set uses validation data from all accident periods, its results will hence correspond to the SLP results.

Based on the different partition strategies in Table \ref{OldParDetails}, we have seventeen ADLP ensembles in total. We use $\text{ADLP}_{i}$ to denote the ADLP ensemble constructed from the $i^{th}$ partition strategy. The fitting algorithm for ADLP is similar to the algorithm for fitting SLP ensembles described in Section \ref{impleLP}, except for Step 4, where the combination weights are estimated using \eqref{EqnOptimalSubTwo} instead of \eqref{EqnOptimalSubOne}. 

\begin{remark} \label{CrossValidationRemark}

The partition of training and validation data is not new in the aggregate loss reserving context. In \cite*{LindRiFeHe20} and \cite*{GaRoMa20}, the data partition is performed on the individual claims level. The loss triangle for models training is constructed by aggregating half of the individual claims, and the validation triangle is built by grouping the other half of the individual claims. This is to accommodate the fact that only aggregate reserving triangles with size 12 x 12 are available in the original study \citep*{GaRoMa20}, which makes partitioning based on the aggregate level difficult due to data scarcity issue. In such scenario, this approach is advantageous as it efficiently uses the limited loss data by using the whole triangle constructed on the validation set to assess the model. Since a relatively large reserving triangle (i.e., 40 x 40) is available in this study and we are more interested on the component models' performance on the latest calendar periods to allocate combination weights based on individual projection accuracy, we decide to directly partition the loss triangle on the aggregate level.

Another approach for partitioning the aggregate loss reserving data is the rolling origin method \citep*{BaRi20, AMAvTaWo22}, which does not rely on the availability of individual claims information. However, this approach is mainly designed for selecting candidate models based on their projection accuracy rather than model combination. 

The data partition strategy proposed in this paper is novel in three different ways. Firstly, it is dedicated to the combination of loss reserving models, as it explicitly considers the difference in component models' performance in different accident periods by partitioning the data by claims maturity. The additional partition step is generally unnecessary for model selection and has not been undertaken in the aforementioned loss reserving literature to our knowledge. However, as discussed in Section \ref{LogScoreOW} and further illustrated in Section \ref{LPDiffPS}, this step is critical for combining loss reserving models as it allows the ensemble to effectively utilise the strengths of component models in different accident periods. Secondly, the proposed partition strategy takes into account the combination of effects from accident periods and development periods in the triangular data format by bringing forward the validation data in older accident periods to train the combination weights in subsequent accident periods. Thirdly, this paper is also the first one that investigates the impact of various ways of splitting the validation data on the ensemble's performance in the aggregate loss reserving context. This paper has found that the ensemble's performance can be sensitive to the choice of split point, which has not been explicitly addressed in the current literature. Motivated by the findings above, the optimal point to split the validation data (see detailed illustrations in Section \ref{LPDiffPS}) has been proposed in this paper, which might provide useful guidance for implementation.

\end{remark}

\begin{remark} \label{Threesubsets}

This paper focuses on the case when $K=2$ (i.e.partitioning the data into two subsets). However, one can divide the data into more subsets and train the corresponding combination weights by simply using $K>2$ in \eqref{EqnOptimalSubTwo}. To illustrate this idea, we provide an example when $K=3$. For instance, for $\text{ADLP}_2$, another split point, say accident period 15, can be introduced after the first split at the accident period 5. We denote the resulting ensemble as $\text{ADLP}_2^{+}$. Similarly, the third subset can be introduced to other ADLP ensembles. Table \ref{ThreeParDetails} shows some examples of the three-subsets partition strategies based on the current ADLP ensembles.

\begin{table}[htb]
\centering \footnotesize
\caption{Details of Three-Subsets Data Partition Strategies \label{ThreeParDetails}}
\begin{tabular}{llll}
Ensembles    & Accident Periods in Subset 1  & Accident Periods in Subset 2   & Accident Periods in Subset 3    \\
$\text{ADLP}_2^{+}$     & 2-5  & 6-15    & 16-40    \\
$\text{ADLP}_8^{+}$     & 2-15  & 16-29     & 30-40   \\
$\text{ADLP}_{10}^{+}$  & 2-17  & 18-31   &  32-40  \\
$\text{ADLP}_{13}^{+}$  & 2-23 &  24-33   &  34-40  \\
\end{tabular}
\end{table}

The details of the performance analysis of the $\text{ADLP}^{+}$ ensembles listed in Table \ref{ThreeParDetails} is given in Appendix \ref{ADLPThreeSubsets}. Overall, there is no significant difference between the out-of-sample performance of the ADLP ensembles and the corresponding $\text{ADLP}^{+}$ ensembles. At least with the data set used in this study, we do not find enough evidence to support the improvement over the current ADLP ensembles by introducing more subsets for training the combination weights. This phenomenon might be explained by the small difference in the performance of component models after the accident period 20, limiting the ensemble's potential benefit gained from the diversity among the performance of component models \citep*{Br96}. Therefore, adding additional subsets in the later accident periods seems redundant at least in this example.

\end{remark}

\subsection{Relationship to the stacked regression approach} \label{StackedregressionDiscuss}

Another popular ensemble scheme is the stacked regression approach proposed by \citet*{Br96}, which uses the Mean Square Error (MSE) as combination criterion. The idea of stacked regression approach has been applied in the mortality forecasts literature. For detailed implementation, one may refer to \cite*{KeShViZi22} and \cite*{Li22}. In \citet*{Lind22}, the authors propose averaging the predictions from an ensemble of neural networks with random initialisation and training data manipulation to obtain the final mortality forecast, which could be thought of as a special case of stacked regression approach where an equal weight is applied to each component model. 

In summary, under the stacked regression approach, the weights allocated to component models are determined by minimising the MSE achieved by the combined forecast: 
\begin{eqnarray} \label{EqnStacked}
\hat{\textbf{w}} &=& \underset{\textbf{w}}{\mathrm{arg\,max}}\; \frac{1}{|D^{val}|}\sum_{y_{ij} \in D^{val}} \left(\sum_{m=1}^{M} w_{m} \hat{\mu}^m_{ij} - y_{ij} \right)^2,
\end{eqnarray}
where $\hat{\mu}^m_{ij}$ is the predicted mean from component model $m$. 

As \citet*{YaVeSiGe18} suggest, both stacked regression and linear pools correspond to the idea of ``stacking''. When MSE is used as optimisation criterion \citep*[i.e., the stacked regression approach proposed by][]{Br96}, the resulting ensemble is a stacking of means. When using the Log Score, the corresponding ensemble is a stacking of predictive distributions.

Although both the stacked regression and linear pools are stacking methods, their objectives are different: the stacked regression seeks to optimise the ensemble's central forecast accuracy, whereas linear pools strive to optimise the calibration and sharpness of distributional forecast. Despite showing success in combining point forecasts, stacked regression is not recommended to combine distributional forecasts \citep*{YaVeSiGe18, KnKr22}. The reasons are twofold as explained below.

Firstly, the combination criterion under the stacked regression approach, which is the Mean Squared Error, is not a strictly proper scoring rule \citep*{GnRa07,KnKr22}. As discussed in Section \ref{modEval}, a strictly proper scoring rule guarantees that the score assigned to the predictive distribution is equal to the score given to the true distribution if and only if the predictive distribution is the same as the true distribution. However, since the MSE only concerns the first moment, a predictive distribution with misspecified higher moments could potentially receive the same score as the true distribution. Therefore, using MSE as the evaluation criterion does not give forecasters the incentives to correctly model higher moments. For an illustrative example, one might refer to Appendix \ref{Mean-var-tradeoff}. 
In contrast, the Log Score is a strictly proper scoring rule, which evaluates the model's performance over the entire distribution, and encourages the forecaster to model the whole distribution correctly \citep*{GnRa07}. This property is particularly desirable for loss reserving, where the model's performance over the higher quantile of the loss distribution is also of interest to insurers and regulators.

The second advantage of using the linear pooling scheme over the stacked regression approach in combining distributional forecasts comes from the potential trade-off between the ensemble's performance on the mean and variance forecasts. In short, minimising the MSE will always improve the mean forecasts, but might harm the ensemble's variance forecasts when the individual variances are unbiased or over-estimated \cite*{KnKr22}. An illustrative proof of this idea can be found in Appendix \ref{Mean-var-tradeoff}. To balance the trade-off between mean and variance forecasts, the combination criterion should take into account the whole distribution (or at least the first two moments). This can be achieved by using a strictly proper scoring rule (e.g., the Log Score).

\section{Implementation of the SLP and ADLP} \label{S_implem}

This section is dedicated to the implementation of the SLP and ADLP framework. It provides  theoretical details for such implementation, and it complements the codes available at \url{https://github.com/agi-lab/reserving-ensemble}.

\subsection{Minorization-Maximization strategy for Log Score optimisation} \label{MMStrategy}

To solve the optimisation problem specified in \eqref{EqnOptimalSubOne} and \eqref{EqnOptimalSubTwo}, we implement the Minorization-Maximization strategy proposed by \citet*{CoDeGi15} which is a popular and convenient strategy for maximizing the Log Score. Instead of directly maximizing the Log Score, a surrogate objective function is maximised. Define
\begin{eqnarray} \label{EqnSurrogateFu}
\phi_{\lambda}(w_m^k,a) &=& \frac{1}{|D_{val}^{k*}|} \sum_{y_{ij} \in D_{val}^{k*}} \sum_{m=1}^{M}\frac{\hat{f}_m(y_{i,j})a_{m}}{\sum_{l=1}^{M}\hat{f}_l(y_{i,j})a_{l}} \ln\left(\frac{w_m^k}{a_{m}}\sum_{l=1}^{M}\hat{f}_l(y_{i,j})a_{l}\right)-\lambda\left(\sum_{m=1}^{M}w_m^k\right),
\end{eqnarray}
where $a_m$ is an arbitrary weight, $\lambda$ is a Lagrange Multiplier, and where $D_{val}^{k*}=D_{val}^{1} \cup...\cup D_{val}^{k}$. Note that for the SLP, we have $w_m^1=\ldots=w_m^k\equiv w_m$. The weights are then updated iterativily by maximizing \eqref{EqnSurrogateFu}:
\begin{eqnarray}
(w^{k}_m)_{i+1} &=& \underset{w}{\mathrm{arg\,max}}\;
\phi_{\lambda}(w^k_m,(w^k_m)_{i}).
\end{eqnarray}
By setting 
$$\frac{\partial \phi_{\lambda}(w^k_m,(w^k_m)_{i})}{\partial w^k_m}=0,$$
we have 
$$w_m^{k}=\frac{1}{\lambda} \frac{1}{|D_{val}^{k*}|} \sum_{y_{ij} \in D_{val}^{k*}} \frac{\hat{f}_m(y_{i,j}) (w^k_m)_{i}}{\sum_{l=1}^{M}\hat{f}_l(y_{i,j})(w^k_l)_{i}}.$$
Now, using the constraint $ \sum_{m=1}^{M} w^k_m=1$ yields
$$\sum_{m=1}^{M} w^k_m=\frac{1}{\lambda} \frac{1}{|D_{val}^{k*}|} \sum_{y_{ij} \in D_{val}^{k*}} \frac{\sum_{m=1}^{M} \hat{f}_m(y_{i,j}) (w^k_m)_{i}}{\sum_{l=1}^{M}\hat{f}_l(y_{i,j})(w^k_l)_{i}}=\frac{1}{\lambda} \frac{1}{|D_{val}^{k*}|} \sum_{y_{ij} \in D_{val}^{k*}} 1=\frac{1}{\lambda} \cdot \frac{|D_{val}^{k*}|}{|D_{val}^{k*}|}=\frac{1}{\lambda}=1.$$
Therefore, $\lambda=1$, and the updated weights in each iteration become: 
\begin{eqnarray}
(w_m^k)_{i+1} &=& (w_m^k)_{i}\sum_{y_{ij} \in D_{val}^{k*}}\frac{\hat{f}_m(y_{i,j})}{\sum_{l=1}^{M}\hat{f}_l(y_{i,j})(w_l^k)_{i}},
\end{eqnarray}
where the weights will be initialised as $(w_m^k)_0=1/M$ such that the constraints $\sum_{m=1}^{M}w_m^k=1$ and $w_m^k \ge 0$ can be automatically satisfied in each iteration. The iterations of weights are expected to converge due to the monotonically increasing property of the surrogate function \citep*{CoDeGi15}. To promote computational efficiency, the algorithm is usually terminated when the difference between the resulting Log Scores from the two successive iterates is less than a tolerable error $\epsilon$ \citep*{CoDeGi15}. We set $\epsilon=10^{-16}$, which is a common tolerable error used in the literature. An outline for the Minorization-Maximization Strategy is given in Algorithm \ref{alg:algoMM}.

\begin{algorithm}[H]
\caption{Minorization-Maximisation algorithm for Log Score maximisation}\label{alg:algoMM}
\begin{algorithmic}[1]
   \State Initialise $(w_1^k)_0=(w_2^k)_0=,...,(w_M^k)_0=\frac{1}{M}$
   \State $i \gets 0$
   \State $\overline{\text{LogS}}_0^k \gets \frac{1}{|D_{val}^{k*}|} \sum_{y_{ij} \in D_{val}^{k*}} \ln (\sum_{m=1}^M (w_m^k)_0 \cdot \hat{f}_m(y_{i,j}))$
\While{$i < \text{MaxIters}$}
   \State Calculate the Log Score: $\overline{\text{LogS}}_i^k \gets \frac{1}{|D_{val}^{k*}|} \sum_{y_{ij} \in D_{val}^{k*}} \ln (\sum_{m=1}^M (w_m^k)_{i} \cdot \hat{f}_m(y_{i,j}))$
   \State Update the combination weights: $(w_m^k)_{i+1} = (w_m^k)_{i}\sum_{y_{ij} \in D_{val}^{k*}}\frac{\hat{f}_m(y_{i,j})}{\sum_{l=1}^{M}\hat{f}_l(y_{i,j})(w_l^k)_{i}}$ 
   \State Update the Log Score: $\overline{\text{LogS}}_{i+1}^k \gets \frac{1}{|D_{val}^{k*}|} \sum_{y_{ij} \in D_{val}^{k*}} \ln (\sum_{m=1}^M (w_m^k)_{i+1} \cdot \hat{f}_m(y_{i,j}))$
   \If{$\overline{\text{LogS}}_{i+1}^k-\overline{\text{LogS}}_{i}^k < 10^{-16}$}
        \State break
    \EndIf
    \State $i=i+1$
\EndWhile
\end{algorithmic}
\end{algorithm}

Note that this process can be used to optimise \ref{EqnOptimalProblem}, as this is a special case of the ADLP with $K=1$.

\subsection{Prediction from Ensemble} \label{PredictEnsemble}

Based on \eqref{ensPredict}, the central estimate of incremental claims by the ensemble can be calculated as follows:
\begin{equation} \label{lmaCentralEstimate}
    \mu^{*}_{ij}=\sum_{m=1}^{M} w_m\cdot\mu_{ij}^m
\end{equation} 
To avoid abuse of notation, $w_m$ in this section can be either the weight under SLP or the weight under ADLP. For ADLP ensembles, $w_m$ is set to $w_m^k$ in the calculation below if $y_{ij} \in D^{k}$.

To provide a central estimation for reserve, we simply aggregate all the predicted mean in the out-of-sample data: $\hat{R}^{*}=\sum_{i,j \in D_{out}} \mu^{*}_{ij}$. Since insurers and regulators are also interested in the estimation of $75^{th}$ reserve quantile, simulation is required. The process for this simulation is outlined in Algorithm \ref{alg:algosimulquantilLP}.

\begin{algorithm}[htb]
\caption{Algorithm for simulating reserve quantile from linear pools ensembles}\label{alg:algosimulquantilLP}
\begin{algorithmic}[1]
    \State Simulate $N$ random variables from $\text{Uniform}(0,1)$; Denote $\tilde{U}=(\tilde{U}_{(1)},...,\tilde{U}_{(N)})$ as the vector of $N$ simulated Uniformly distributed random variables
    \State Denote $\tilde{Y}^{*}_{ij}=(\tilde{Y}^{*}_{ij,(1)},...,\tilde{Y}^{*}_{ij,(N)})$ as the vector of $N$ simulated variables for cell $(i,j)$ from the ensemble
    \For{$n=1,..,N$}                    
        \If{$\tilde{U}_{(n)} \in (\sum_{m=1}^{l-1}w_m,\sum_{m=1}^{l}w_m]$}
            \State simulate a random variable from the $l^{th}$ component distribution, denoted as $\tilde{Y}_{ij}^{(l)}$
            \State Set $\tilde{Y}^{*}_{ij,(n)}=\tilde{Y}_{ij}^{(l)}$
        \EndIf 
    \EndFor
    \State Repeat the above two steps for each cell $(i,j)$
    \State Calculate the simulated reserve based on the simulated random variables for each cell $(i,j)$: $\tilde{R}^{*}=(\tilde{R}^{*}_{(1)},\tilde{R}^{*}_{(2)},...,\tilde{R}^{*}_{(N)})=(\sum_{i,j\in D^{out}}\tilde{Y}_{ij,(1)}^{*},\sum_{i,j\in D^{out}}\tilde{Y}_{ij,(2)}^{*},...,\sum_{i,j\in D^{out}}\tilde{Y}_{ij,(N)}^{*})$
    \State Calculate the empirical $75^{th}$ quantile of the $N$ simulated reserves for the ensemble: $R^{*}_{75}$
\end{algorithmic}
\end{algorithm}

\section{Comparison of the predictive performance of the ensembles} \label{comparepred}

\subsection{Measuring distributional forecast accuracy: Log Score} \label{LogS}

To assess and compare the out-of-sample distributional forecast performance of different models, the Log Score is averaged across all the cells in the lower triangle :
\begin{eqnarray}
\text{LogS}^{out} &=& \frac{1}{|D^{out}|}\sum_{i,j \in D^{out}}\ln \left(\widehat{f}(y_{i,j})\right).
\end{eqnarray}

Since loss reserving models tend to perform differently in different accident periods, we also summarise the Log Score by accident periods: 
\begin{eqnarray} \label{LSAPEquation}
\text{LogS}_{\text{AP}_i}^{out} &=& \frac{1}{|D_{\text{AP}_i}^{out}|}\sum_{j \in D_{\text{AP}_i}^{out}}\ln\left(\widehat{f}(y_{i,j})\right),
\end{eqnarray}
where $D_{\text{AP}_i}^{out}$ denotes the set of out-of-sample incremental claims in accident period $i$. 

\rev{
\begin{remark}
    Besides Log Score, another popular proper score in literature is the Continuously Ranked Probability Score (CRPS) \citep*{GnRa11b,OpVaVan17}. This paper (and associated R package) also incorporates the CRPS as an alternative evaluation metric for the methods proposed. Further details and illustrative results are provided in Appendix \ref{Appendix:weightedCRPS}. Both scores generally lead to consistent conclusions. 
\end{remark}
}

\subsection{Statistical tests} \label{StatTestMethod}

To determine whether the difference between the performance of two competing forecasting models is statistically significant, statistical tests are necessary. We consider the Diebold-Mariano test in this paper due to its high flexibility and wide application in probabilistic forecasting literature \citep*{DiMa02,GnKa14}. Diebold-Mariano test is one of the few tests that does not require loss functions to be quadratic and the forecast errors to be normally distributed, which is what we need in our framework. Furthermore, the Diebold-Mariano test is specifically designed to compare forecasts and inherently emphasises out-of-sample performance \citep{Di15}. This feature aligns well with our objectives, as actuaries conducting reserving exercises are primarily concerned with the accuracy of projections. The performance of the Diebold-Mariano test has been discussed in \citet{CoKu11}, who discovered a tendency for the test to favour simpler benchmark models. As a result, the Diebold-Mariano test serves as a conservative criterion for evaluating the performance of our proposed ensemble strategies.

Consider the null hypothesis stating that model $G$ has equal performance as model $F$. The Diebold-Mariano test statistic can then be specified as \citep*{DiMa02} 
\begin{eqnarray} \label{DMTestEquation}
t_n &=& \sqrt n \frac{\overline{S}_n^F-\overline{S}_{n}^G}{\hat{\sigma}_{n}}.
\end{eqnarray}
where $\overline{S}_n^F$ and $\overline{S}_n^G$ are the average scores (here, Log Score) received by models $F$ and $G$ in out-of-sample data, and where $\hat{\sigma}_{n}$ is the estimated standard deviation of the score differential. That is,  \begin{equation}
\hat{\sigma}_{n}=\sqrt{\frac{1}{|D^{out}|}\sum_{i,j \in D^{out}}(\ln(\widehat{f_F}(y_{i,j}))-\ln(\widehat{f_G}(y_{i,j})))^2}.
\end{equation}
Assuming the test statistic to asymptotically follow a standard Normal distribution, the null hypothesis will be rejected at the significance level of $\alpha$ if the test statistic is greater than the $(1-\alpha)$ quantile of the standard Normal distribution.

\subsection{Measuring performance on the aggregate reserve level} \label{MeasureAccuCentralReserve}

On the aggregate level, an indication of the performance of the central estimate for reserve by model $m$ can be provided by the relative reserve bias:
\begin{eqnarray}
R^{\text{bias}} &=& \frac{\hat{R}^{m}-R^{\text{True}}}{R^{\text{True}}}.
\end{eqnarray}
Since insurers and regulators are also interested on the estimation of the $75^{th}$ quantile of reserve, a model's performance in this quantile can be indicated by its relative reserve bias at the $75^{th}$ quantile:
\begin{eqnarray}
R^{\text{bias}}_{75} &=& \frac{\hat{R}^{m}_{75}-R^{\text{True}}_{75}}{R^{\text{True}}_{75}},
\end{eqnarray}
where $R^{\text{True}}_{75}$ is the $75^{th}$ quantile of the true outstanding claims, and $\hat{R}^{m}_{75}$ is the estimated $75^{th}$ quantile by model $m$. The calculation details for $\hat{R}^{m}_{75}$ can be found in Appendix \ref{CalReserQuant}. 

\begin{remark}
Of course, $R^{\text{True}}$ and $R^{\text{True}}_{75}$ are not observable in practice. Thanks to our use of simulated data (see Section \ref{Exampledata}) though, we are able to calculate those from the full simulations and here have been calculated using simulated data in the bottom triangles. Furthermore, as these metrics are not proper scoring rules, care should be taken in usage of these results in ranking models.
\end{remark}

\section{Results and discussion} \label{empiricalResults}

After presenting the ensemble modelling framework, this section provides numerical examples to illustrate the out-of-sample performance of the proposed linear pool ensembles SLP and ADLP, as well as the BMV and EW. Following the introduction of the synthetic loss reserving data in Section \ref{Exampledata}, Section \ref{basicLPEns} and Section \ref{LPDiffPS} analyse the distributional forecast performance of the SLP and the ADLP ensembles on incremental claims level, respectively. Section \ref{StatsTestDFP} analyses the difference in models' performance using statistical test results. The performance of the proposed ensembles is also analysed at the aggregate reserve level in Section \ref{PredPerfAR}. Finally, we share some insights on the properties of component models by analysing the optimal combination weights yielded by our framework.

\subsection{Example data} \label{Exampledata}

Our models are tested using the \texttt{SynthETIC} simulator \citep*{AvTaWaWo21}, from which triangles for aggregate paid loss amounts, reported claim counts, and finalised claim counts are generated and aggregated into 40x40 triangles. To better understand the models' predictive performance and test the consistency of the performance, 100 loss data sets are simulated, and models are fit to each of the 100 loss data sets.

Simulated data is preferred in this study as it allows the models' performance to be assessed over the distribution of outstanding claims, and the quality of distribution forecast is the main focus of this paper. This paper uses the default version of \texttt{SynthETIC}, with the following key characteristics \citep*{AvTaWaWo21}:
\begin{itemize}
    \item Claim payments are long-tailed: In general, the longer the settlement time, the more challenging it is to provide an accurate estimate of the reserve due to more uncertainties involved; \citep*{JaCaFe19}
    \item High volatility of incremental claims;
    \item Payment size depends on claim closure: The future paid loss amount generally varies in line with the finalisation claims count (i.e. the number of claims being settled), particularly in earlier accident periods.  
\end{itemize}
The key assumptions above are inspired by features observed in real data and aim to simulate the complicated issues that actuaries can encounter in practice when modelling outstanding claims \citep*{AvTaWaWo21}. With such complex data patterns, it is challenging for a single model to capture all aspects of the data. Therefore, relying on a single model might increase the risk of missing important patterns present in the data. 

\subsection{Predictive performance at the incremental claims level} \label{PredPerIncreCl}

In terms of distributional forecast accuracy, the linear pool ensembles SLP and ADLP outperform both the EW ensemble and the BMV \textbf{at the incremental claims level}, measured by the average Log Score over 100 simulations. Furthermore when appropriately partitioning the validation data into two subsets, the performance of the ADLP is better than that of the SLP.

\subsubsection{SLP: the Standard Linear Pool ensemble} \label{basicLPEns}

\begin{figure}[htb]
\begin{minipage}{\textwidth}
\begin{minipage}[t]{0.49\textwidth}
    \centering
    \includegraphics[width=\textwidth]{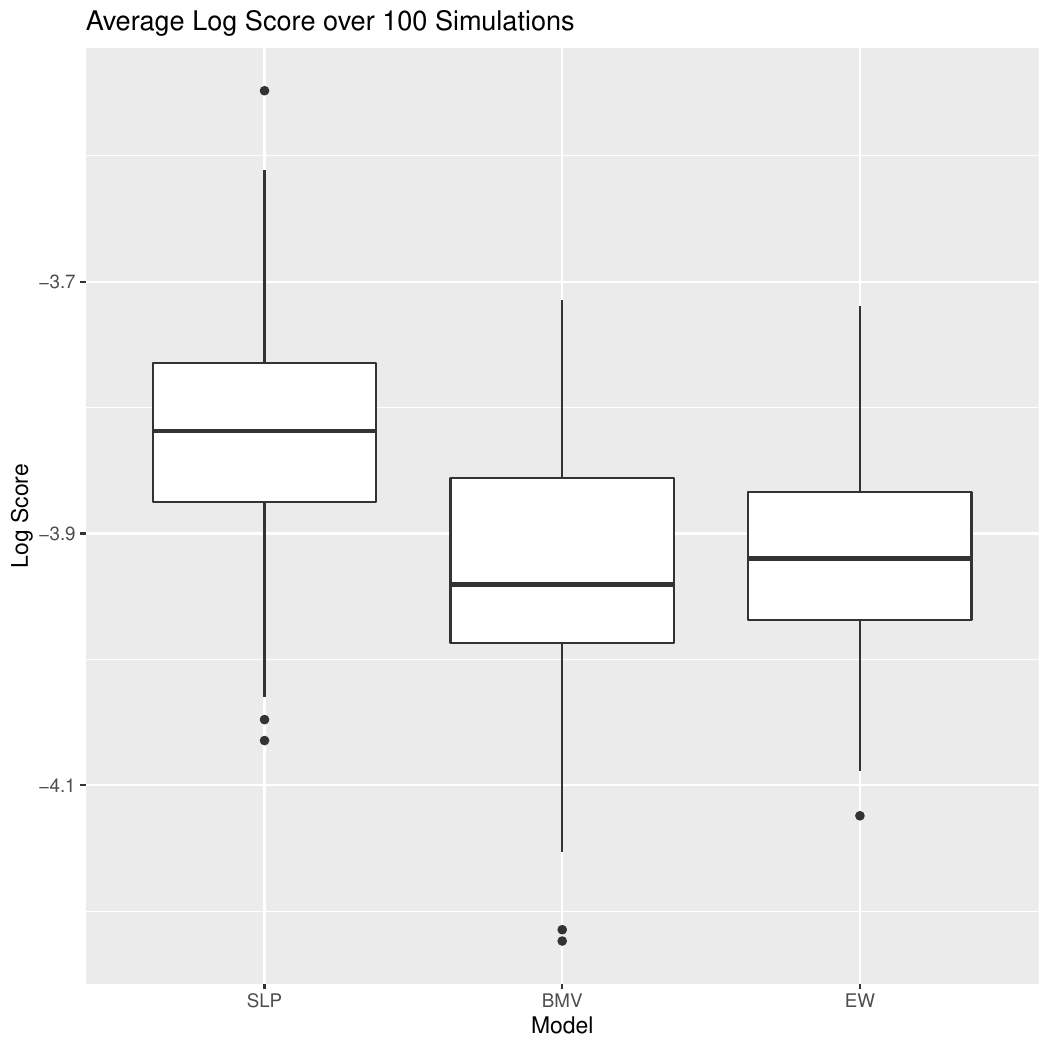}
    \captionof{figure}{Distribution of Log Score over 100 simulations (higher is better): comparison among SLP, BMV and EW}
    \label{LP0COMPLS} 
    \end{minipage}
  \begin{minipage}[t]{0.49\textwidth}
    \centering
    \includegraphics[width=\textwidth]{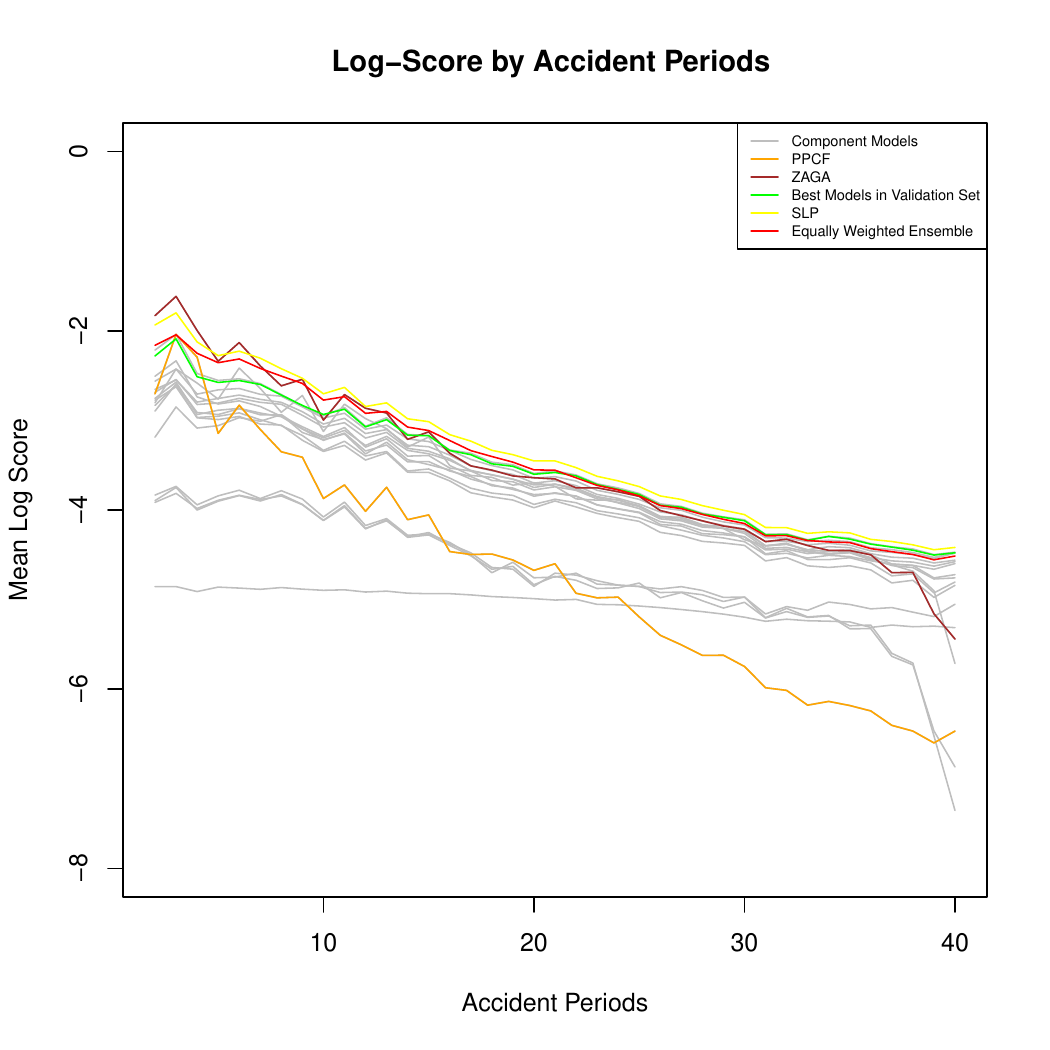}
    \captionof{figure}{Mean Log Score by accident periods (higher is better)}
    \label{LP0COMPLSACC} 
  \end{minipage}
  \hfill
  \end{minipage}
\end{figure}

The distribution of the Log Score received by SLP, BMV, EW (i.e., the equally weighted ensemble) is illustrated in Figure \ref{LP0COMPLS}. Even without any partition of the validation set for training combination weights, the linear pool ensemble achieves a higher average Log Score than its competing strategies (which means it is preferred), with similar variability in performance.

The performance of different methods in each accident period is illustrated in Figure \ref{LP0COMPLSACC}, which shows the Log Score by accident periods being calculated using \eqref{LSAPEquation}, averaging over 100 simulations. The SLP strictly outperforms the BMV in all accident periods. Although the BMV, which is the GAMLSS model with Log-Normal distributional assumption (i.e., $\text{GAMLSS}_{LN}$) in this example, has the highest \textbf{average} Log Score among the component models, it does not yield the best performance in every accident period. For instance, the BMV under-performs the PPCF model in accident periods 3, and it is beaten by ZAGA GLM in earlier accident periods. Therefore, the common model selection strategy, which relies on the BMV to generate predictions for all accident periods, might not be the optimal solution.

Additionally, the SLP strictly outperforms the equally weighted ensemble after accident period 20. The under-performance of the equally weighted ensemble might be explained by the relatively poor performance of several models in late accident periods, which drag down the Log Score attained by the equally weighted ensemble. This relative poor performance is due to the fact that some models are more sensitive to data scarcity in more immature (later) accident years. For instance, the cross-classified models assume a separate parameter for each accident period, and later accident periods have little data available for training. Therefore, despite the simplicity of the equally weighted ensemble, blindly assigning equal weight to all component models tends to ignore the difference in the component models' strengths and might increase the risk of letting a few inferior models distort the ensemble's performance.          

However, there is no substantial improvement over the equally weighted ensemble brought by SLP at the earlier half of the accident periods. The unsatisfactory predictive performance of SLP at earlier accident periods (i.e., mature accident periods) might be explained by the fact that a component model receives the same weight across all accident periods under the SLP, which is what motivated the development of the ADLP in Section \ref{LogScoreOW}. More specifically, and in this example, although the PPCF and the ZAGA model have relatively low overall Log Score due to their poor performance in immature accident periods, they have the best predictive performance among the component models at mature accident periods. However, since the weight allocated to each model reflects its overall predictive performance in all accident periods, the PPCF and the ZAGA GLM are assigned small weights as shown in Section \ref{ModelWeightsInterpret}, regardless of their outstanding predictive performance in earlier accident periods. 
The ADLP offers a solution to this problem, as illustrated in the following section.

\subsubsection{$\text{ADLP}_1$ to $\text{ADLP}_{18}$: Linear Pool ensembles with different data partition strategies} \label{LPDiffPS}

\begin{figure}[htb]
\begin{minipage}{\textwidth}
\begin{minipage}[t]{0.49\textwidth}
   \centering
    \includegraphics[width=\textwidth]{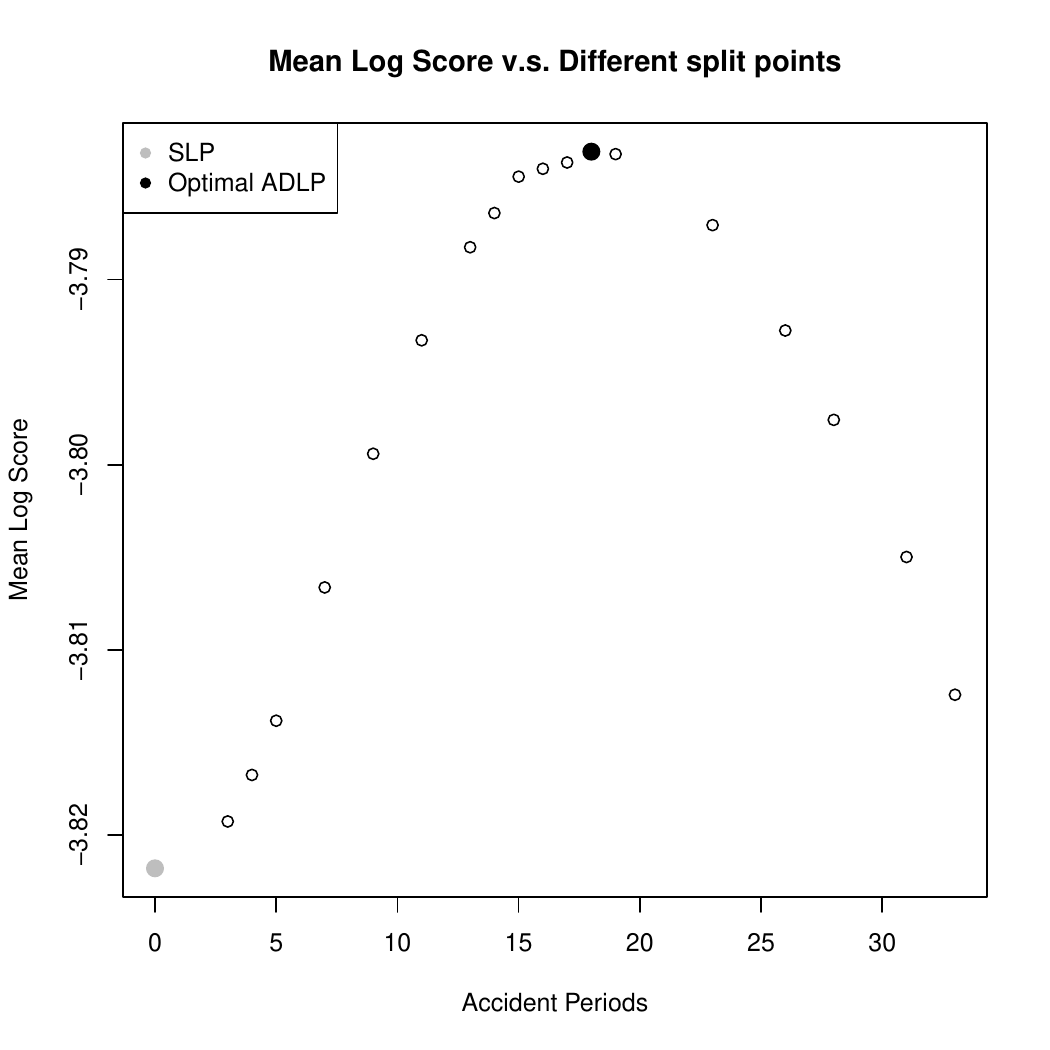}
    \captionof{figure}{Mean Log Score Plot: comparison among different partition strategies  (higher is better)}
    \label{LSMulti} 
    \end{minipage}
  \begin{minipage}[t]{0.49\textwidth}
    \centering
    \includegraphics[width=\textwidth]{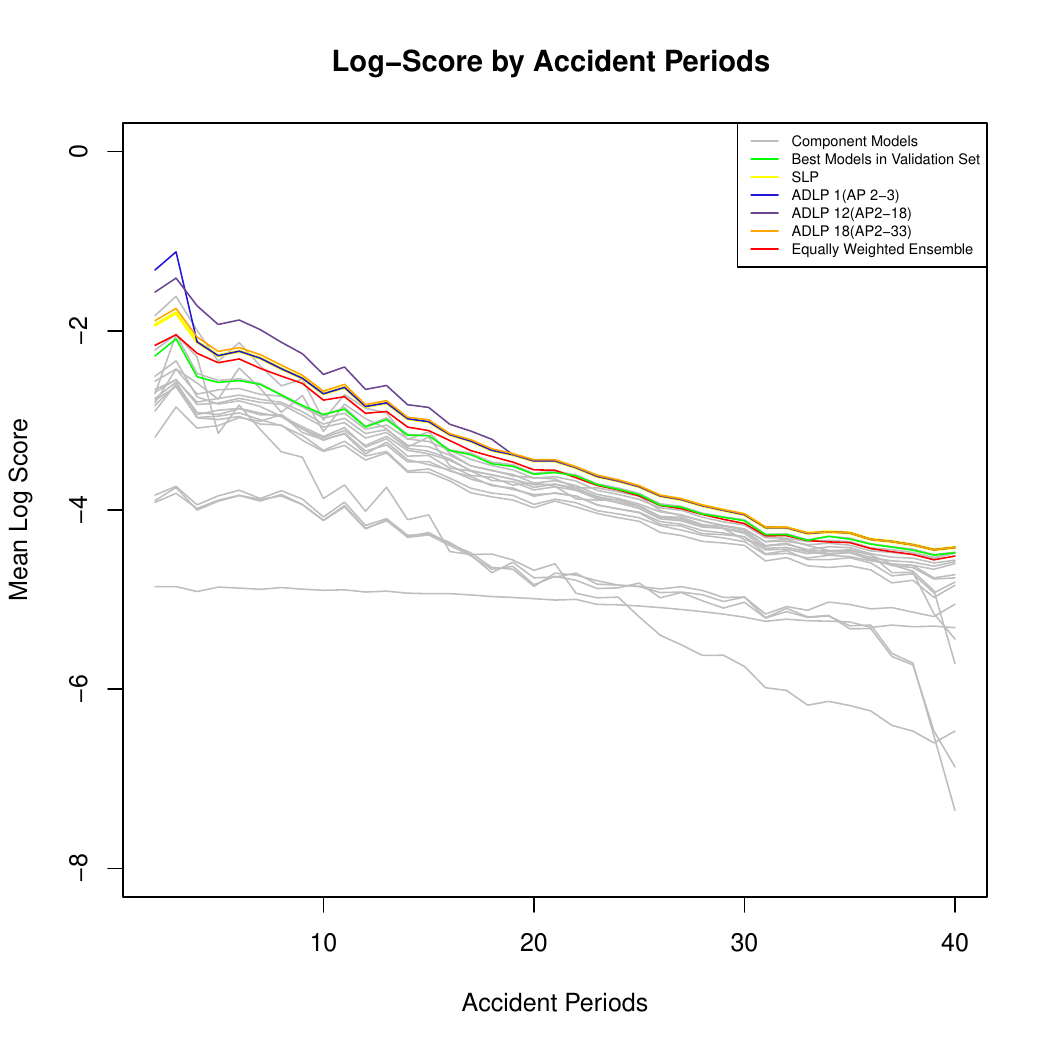}
    \captionof{figure}{Log Score Plot by Accident Periods (higher is better): comparison among different partition strategies}
    \label{LSAPMulti} 
  \end{minipage}
  \hfill
  \end{minipage}
\end{figure}

Figure \ref{LSMulti} plots the \textbf{out-of-sample} Log Score attained by ADLP ensembles with the different split points specified in Table \ref{OldParDetails}, averaging over 100 simulations. The grey point represents the mean Log Score attained by SLP, which is a special case of ADLP ensemble with split point at accident period 0 (i.e., there is no subset). All the ADLP ensembles outperform SLP based on Log Score, supporting the advantages of data partitioning as discussed in previous sections.  

As per Figure \ref{LSMulti}, the mean Log Score forms an inverse U shape, with the peak when splitting at accident period 18. The shape of the mean Log Score can be explained by the following reasons. If the split point is too early (e.g., split point at accident period 3), there is little improvement brought by ADLP over SLP due to the small number of data points in the first subset. This idea can also be illustrated in Figure \ref{LSAPMulti}, which dissects the ensembles' performance into accident periods. Although $\text{ADLP}_1$, represented by the orange line in Figure \ref{LSAPMulti}, has the best performance in the second and third accident period, its Log Score falls to the SLP's level afterward. Therefore, the overall difference between $\text{ADLP}_1$ and SLP is relatively tiny. As more accident periods are incorporated into the first subset, there is a greater overall difference between ADLP ensembles and SLP, which explains the increasing trend of mean Log Score before accident period 18.

However, for late split points (e.g., the split point at accident period 33), the performance of ADLP ensembles also tends to SLP as shown in Figure \ref{LSAPMulti}. This is because the first subset of those ADLP ensembles with late split points covers most of the accident periods,  and the SLP can be interpreted as a special case of ADLP ensembles with one set that includes all accident periods. Additionally, late split points might make it hard for the ensemble to focus on and thus take advantage of the high-performing models in earlier accident periods.

Based on the above reasoning, an ideal partition strategy should balance the ensemble's performance in both earlier and later accident periods. In this study, the optimal split point is attained at accident period 18. As per Figure \ref{LSAPMulti}, $\text{ADLP}_{12}$, which has a split point on accident period 18, substantially outperforms its competitors in the first fifteen accident periods by effectively capturing the high Log Score attained by a few outstanding models.

However, we want to acknowledge that this example is for illustration purposes. The recommendation made above may be more relevant to the loss data sets that share similar characteristics with the synthetic data used in this paper. For practical application, if there is no substantial difference among the performance of component models in earlier accident periods, one may consider simply allocating the first half of accident periods to the first subset.

\subsubsection{Analysing the difference in distributional forecast performance using statistical tests} \label{StatsTestDFP}

The Log Scores of the ADLP in Figure \ref{LSMulti} are clearly higher than that of the SLP, but we haven't established whether this difference is statistically significant yet. To analyse the statistical significance of the improvement over the traditional approaches brought by the ADLP ensembles, we run the Diebold-Mariano test on each simulated dataset (see \eqref{DMTestEquation}), with the null hypothesis stating that the two strategies have equal performance. In particular, we test $\text{ADLP}_{12}$, which has the best predictive performance measured by Log Score, against the EW and BMV, respectively.
If the null hypothesis is rejected, the $\text{ADLP}_{12}$ should be favoured. 

\begin{figure}[htb]
\begin{minipage}{\textwidth}
\begin{minipage}[b]{0.33\textwidth}
    \centering
\includegraphics[width=\textwidth]{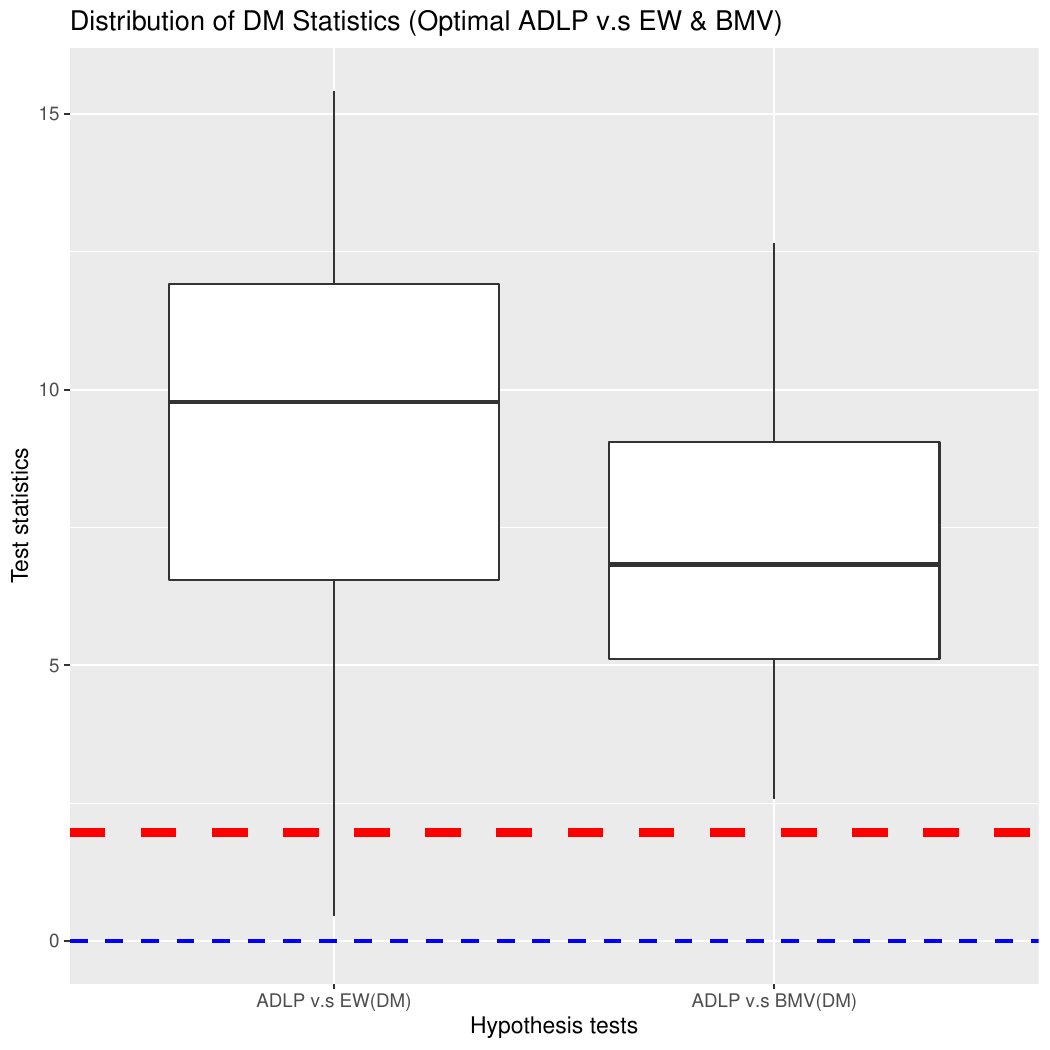}
    \caption{Test Statistics: $\text{ADLP}_{12}$ v.s EW and BMV}
    \label{TestStatslLP0EWBMV}
    \end{minipage}
\begin{minipage}[b]{0.33\textwidth}
    \centering
    \includegraphics[width=\textwidth]{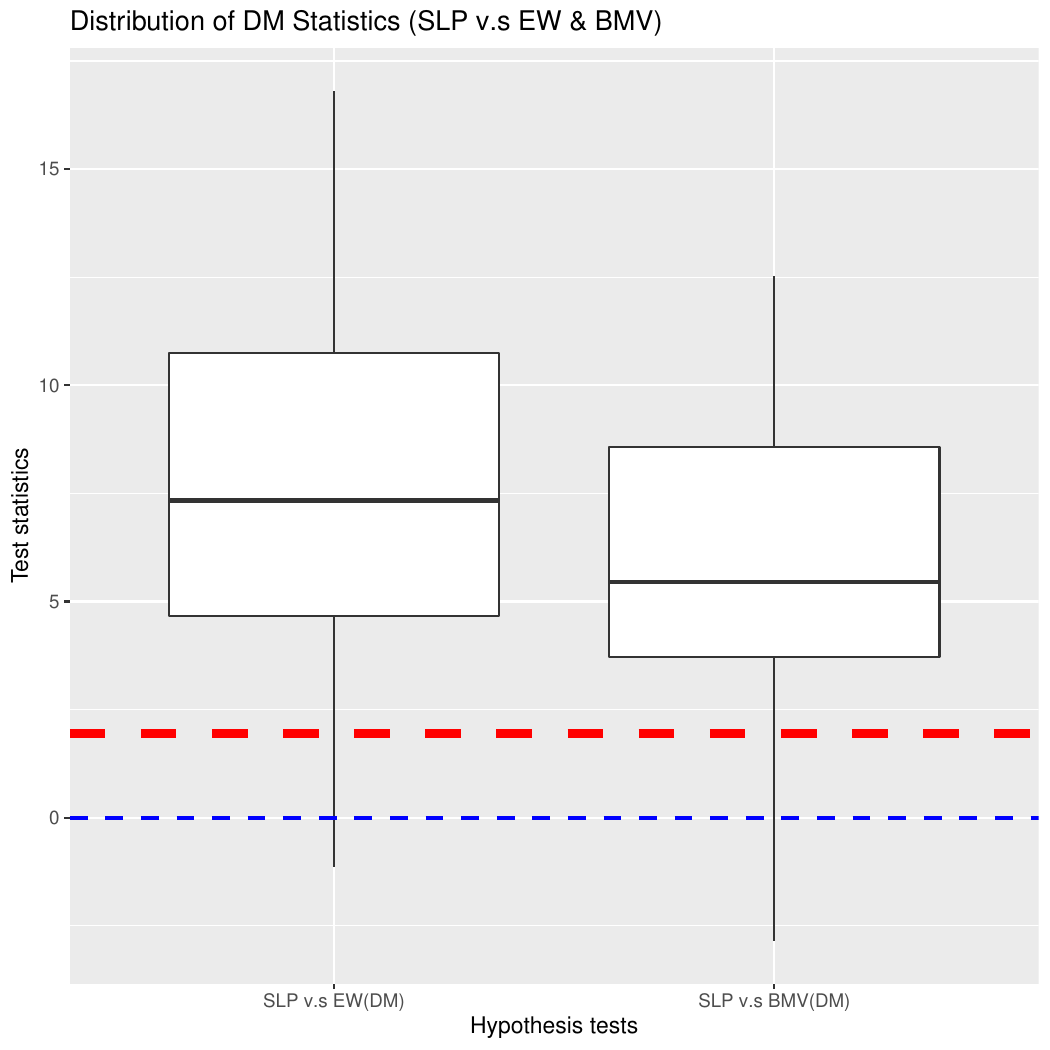}
    \captionof{figure}{Test Statistics: SLP v.s EW and BMV}
    \label{TestStatslSLPEWBMV} 
    \end{minipage}
  \begin{minipage}[b]{0.33\textwidth}
    \centering
    \includegraphics[width=\textwidth]{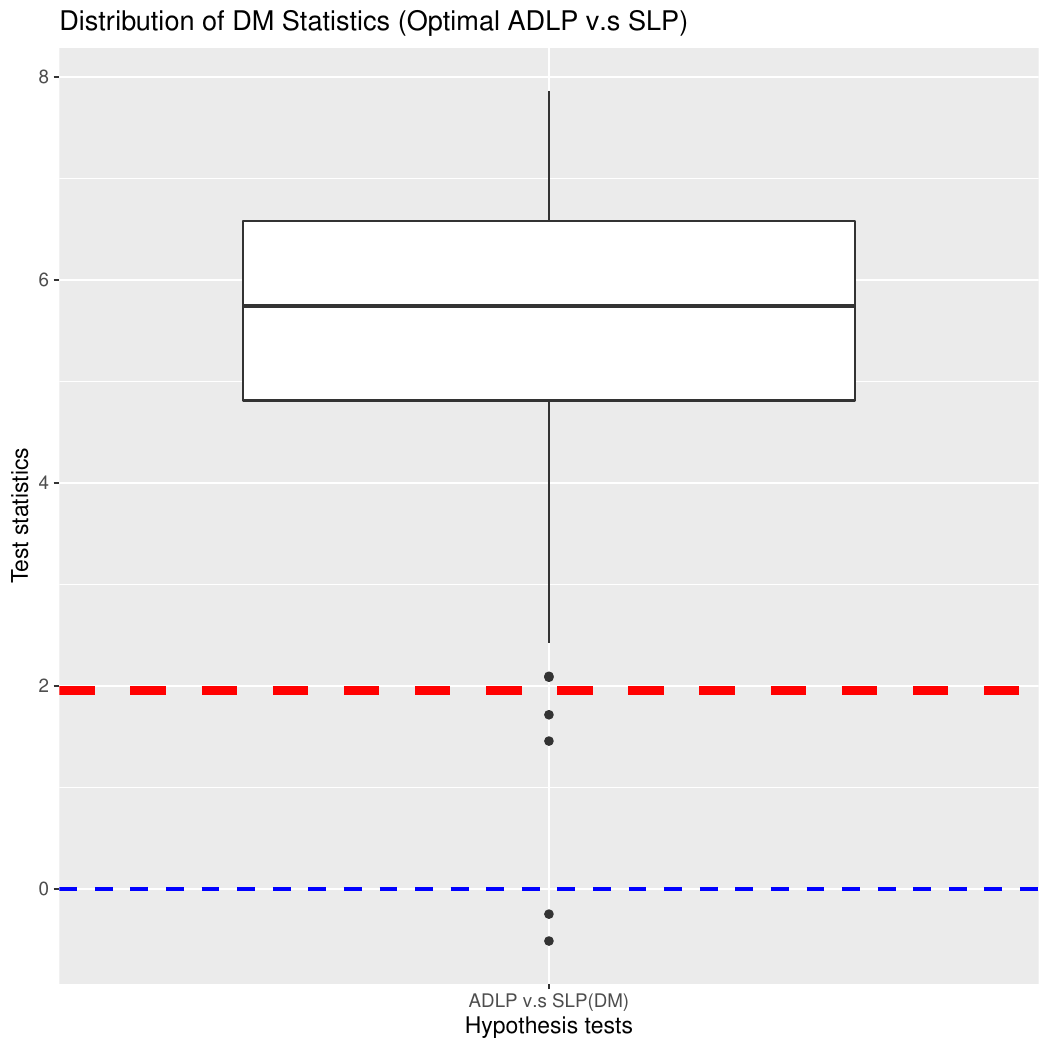}
    \captionof{figure}{Test Statistics: $\text{ADLP}_{12}$} v.s SLP
    \label{TestStatslADLP8vsSLP} 
  \end{minipage}
  \hfill
  \end{minipage}
 \end{figure}

Figure \ref{TestStatslLP0EWBMV} illustrates the distribution of test statistics under the Diebold-Mariano test, with the red dashed line marking the critical value at the customary $5\%$ significance level. The test statistics above the red dashed line indicate those datasets where the null hypothesis has been rejected, and the number of rejections under each test is summarised in Table \ref{NumDatasetsADLP8}. Since the null hypothesis is rejected for most simulations and almost all the test statistics are above zero (as marked by the blue dashed line),the DM test implies a decision in favour of $\text{ADLP}_{12}$.

\begin{table}[htb]
\centering
\caption{The Number of Datasets (out of 100) when $H_0$ is rejected under $5\%$ significance level \label{NumDatasetsADLP8}}
\begin{tabular}{llll}
 Hypothesis Tests                          & DM Test    \\ \hline
$\text{ADLP}_{12}$ vs EW             & 98           \\
$\text{ADLP}_{12}$ vs BMV            & 100   \\
SLP vs EW                                  & 91          \\
SLP vs BMV                                 & 91          \\ 
$\text{ADLP}_{12}$ vs SLP            & 96            \\  \hline
\end{tabular}
\end{table}

We now investigate where the difference in the performance between $\text{ADLP}_{12}$ and BMV or the EW comes from. Since we expect the improvement over the traditional approaches to partially come from the ability of linear pools to optimise the combination of component models based on Log Score, firstly, we test the Standard Linear Pool (SLP) against the traditional approaches for each simulated data set. As per Figure \ref{TestStatslSLPEWBMV} and Table \ref{NumDatasetsADLP8}, the null hypothesis is rejected for most simulated data sets, and most of the test statistics are positive, SLP is favoured over BMV and the EW in most cases.

As per the discussion in Section \ref{LogScoreOW}, we expect $\text{ADLP}_{12}$ to improve SLP by better taking into the typical reserving characteristics found in the simulated data set. 
Based on the results shown in Table \ref{NumDatasetsADLP8} and Figure \ref{TestStatslSLPEWBMV}, $\text{ADLP}_{12}$ is favoured by the statistical tests in most simulated data sets.  

Therefore, the outstanding performance of $\text{ADLP}_{12}$ can be decomposed into two parts. Firstly, it improves the performance of the model selection strategy and the equally weighted ensemble, which can be credited to the advantages of linear pools in model combination. Secondly, as a linear pool ensemble being specially tailored to the general insurance loss reserving purpose, $\text{ADLP}_{12}$ further improves the performance of SLP by considering the impact from accident periods and development periods on the ensemble's performance.

\subsection{Predictive performance on aggregate reserve level} \label{PredPerfAR}

\begin{figure}[H]
\centering
\begin{minipage}{\textwidth}
\begin{minipage}[t]{0.49\textwidth}
    \centering
    \includegraphics[width=\textwidth]{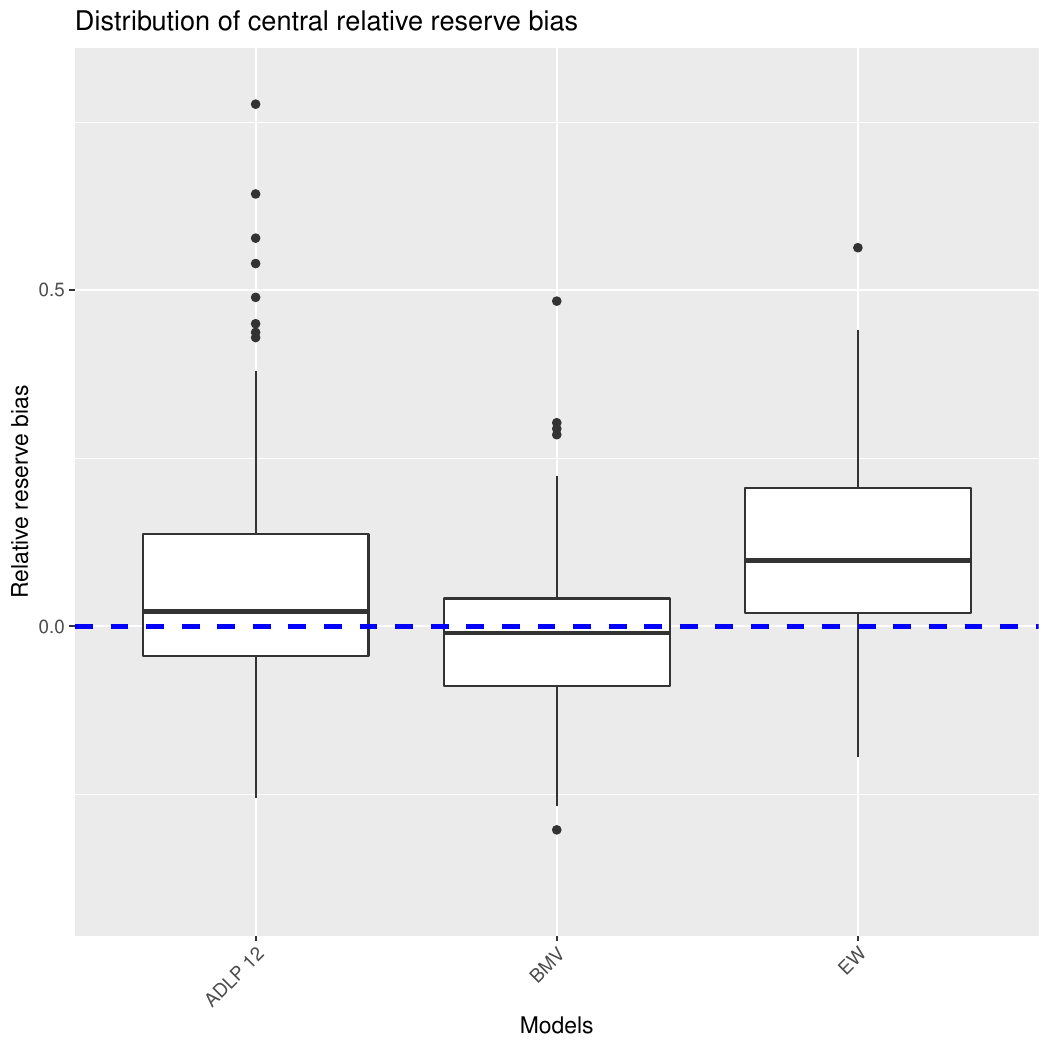}
    \captionof{figure}{Distribution of central relative reserve bias for $\text{ADLP}_{12}$, BMV and EW}
    \label{boxplotCentral} 
    \end{minipage}
  \begin{minipage}[t]{0.49\textwidth}
    \centering
    \includegraphics[width=\textwidth]{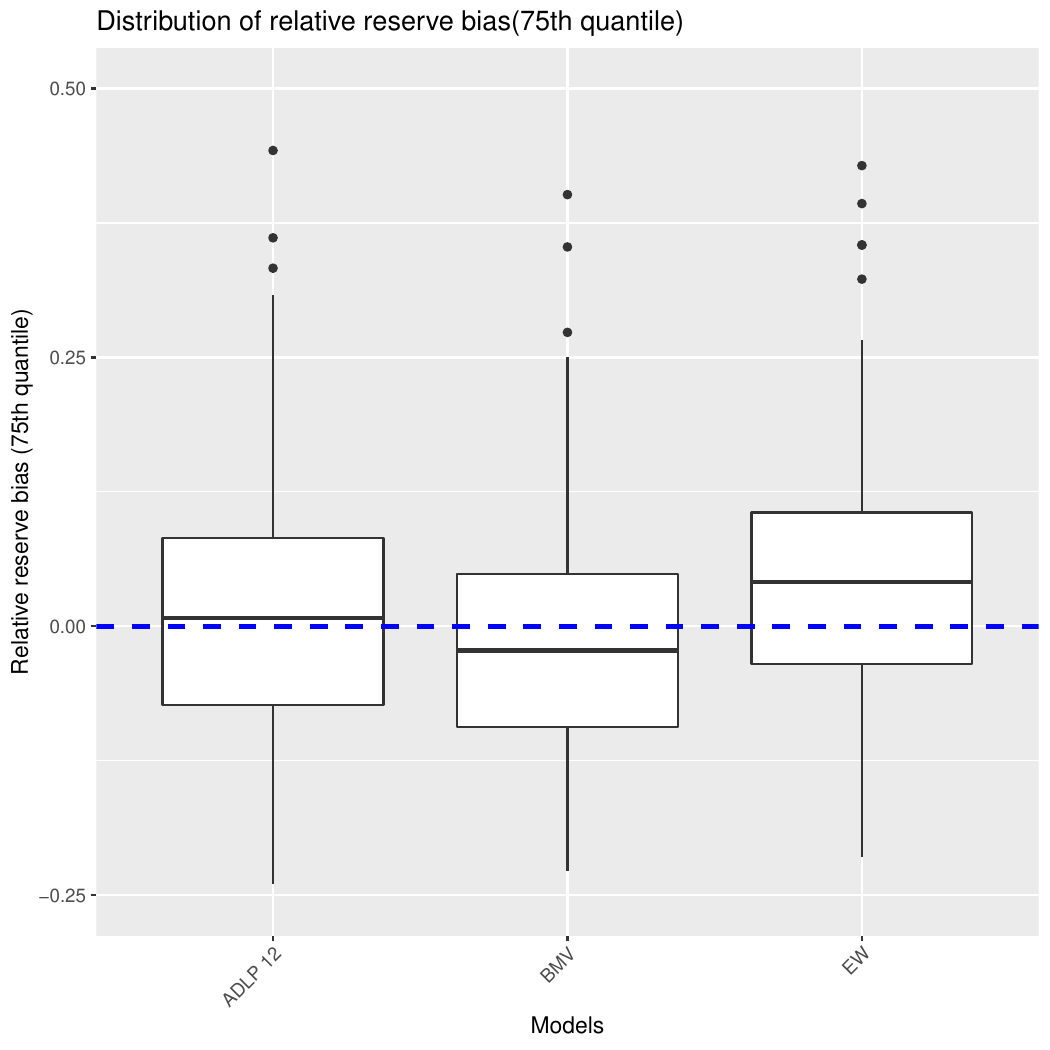}
    \captionof{figure}{Distribution of relative reserve bias at $75^{th}$ quantile for $\text{ADLP}_{12}$, BMV and EW}
    \label{boxplot75Quantile} 
  \end{minipage}
  \hfill
  \end{minipage}
 \end{figure}

Since the central estimation of reserve and its corresponding risk margin at the $75^{th}$ quantile are required to be reported by insurers under Australian prudential regulatory standards \citep{gps340}, we took this as an example of actuarial application and have also calculated those two quantities following the methodology specified in Section \ref{MeasureAccuCentralReserve} in order to gain some insights regarding the performance of the three different models (ADLP, EW, and BMV). The box plots in Figure \ref{boxplotCentral} and Figure \ref{boxplot75Quantile} illustrate the distributions of relative reserve bias for central and quantile estimations, respectively, across 100 simulations. A blue dashed line indicates the point of $0\%$ bias in each plot.

The equally weighted ensemble displays a notably significant inclination towards overestimation, evident under both central and quantile estimation approaches. Although the BMV demonstrates relatively good performance in central reserve estimation, it frequently under-predicts at the $75^{th}$ quantile. In contrast, the ADLP ensemble experiences fewer instances of underestimation or overestimation biases under both central and quantile reserve estimation methods.
\begin{remark} Note that, as discussed in Section \ref{MeasureAccuCentralReserve}, the reserve bias metrics are not proper scoring rules, and so care needs to be taken in interpreting the above results. Nevertheless, we consider the above useful in terms of developing intuition regarding the performance of each model. \end{remark}

\subsection{Analysis of optimal model weights} \label{ModelWeightsInterpret}

\begin{figure}[htb]
\centering
\begin{minipage}{\textwidth}
\begin{minipage}[t]{0.49\textwidth}
    \centering
    \includegraphics[width=\textwidth]{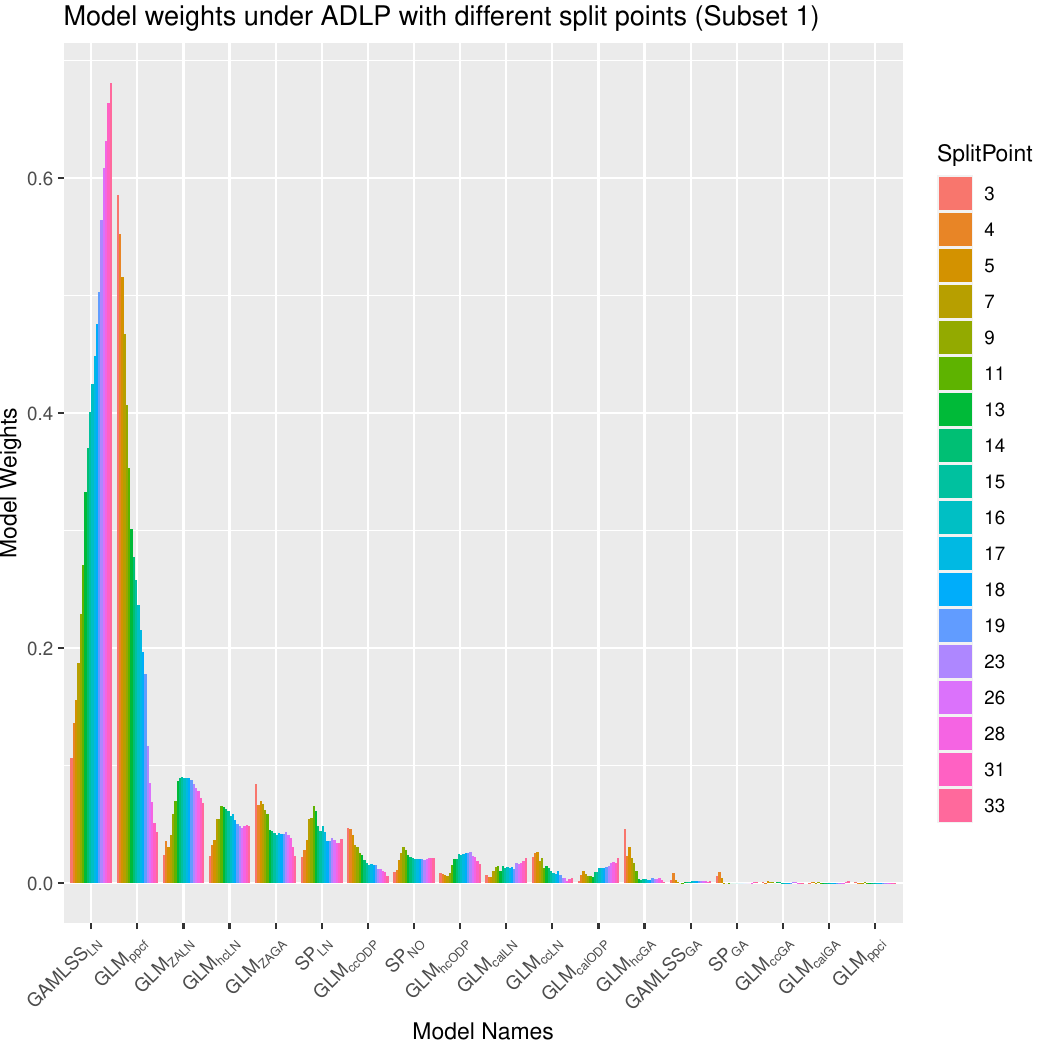}
    \captionof{figure}{Combination Weights in Subset 1: ADLP ensembles}
    \label{WeightsADLPSplitS}

    \end{minipage}
  \begin{minipage}[t]{0.49\textwidth}
    \centering
    \includegraphics[width=\textwidth]{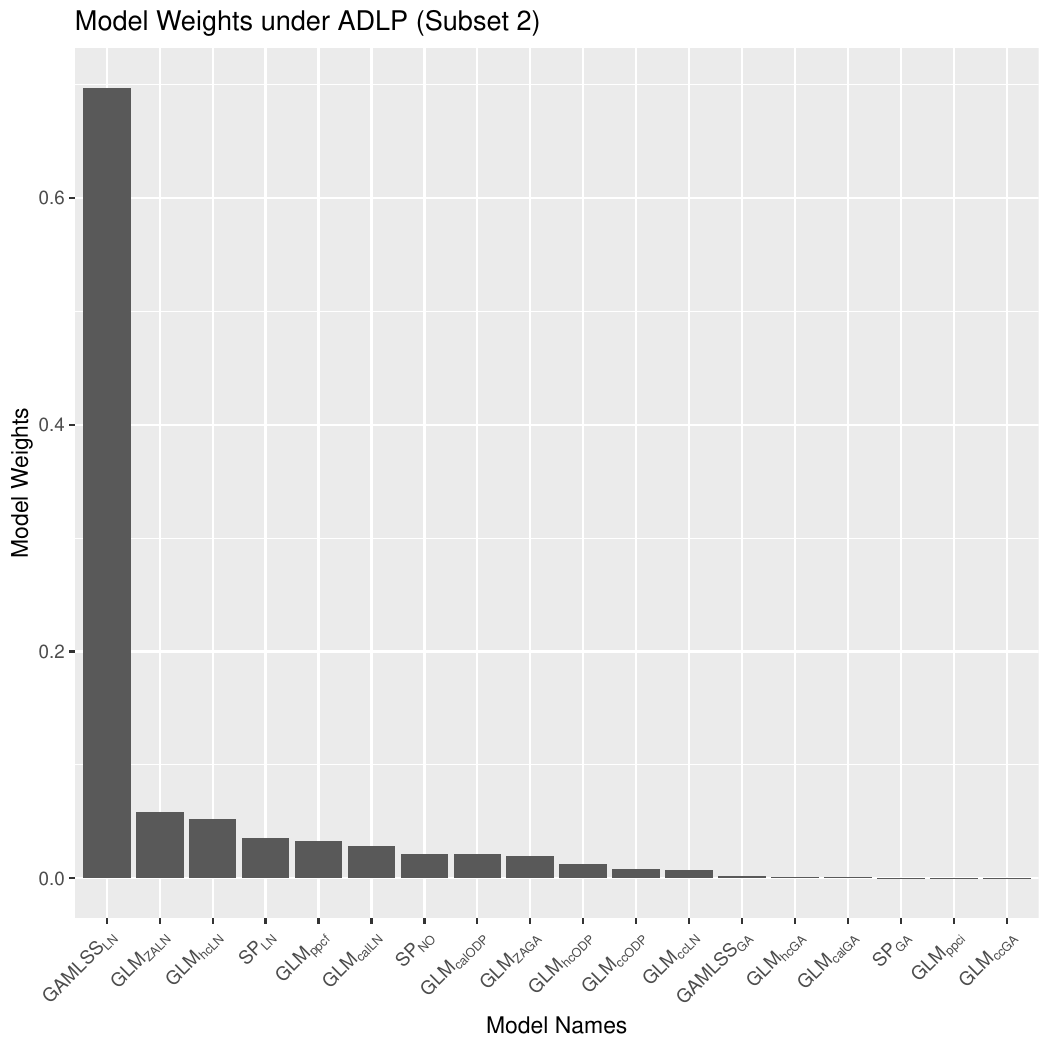}
    \captionof{figure}{Combination Weights in SLP and Subset 2 of ADLP}
    \label{WeightsADLPSplitS2}
    \end{minipage}
  \hfill
  \end{minipage}
 \end{figure}

Analyzing the combination weights can potentially help modelers gain insights about the component models' properties, and examples can be found in \citet*[Chapter 5 of][pp. 151--165]{Tay00} and \citet*{KeShViZi22}. Therefore, we extract the combination weights from ADLPs, with Figure \ref{WeightsADLPSplitS} plotting the model weights in subset 1 using different split points, and Figure \ref{WeightsADLPSplitS2} illustrating the model weights in subset 2. Since model weights under ADLPs in the second subset are trained using the whole validation data---remember \eqref{EqnOptimalSubTwo}, they will be identical to the combination weights under the SLP.

Several models are worth highlighting. The weight allocated to $\text{GAMLSS}_{LN}$, which is the BMV (identified as the model with the highest Log Score \textbf{averaging} over different accident periods) in this example, is increasing with the split point as shown in Figure \ref{WeightsADLPSplitS}. Since the combination weights reflect the relative validation performance of component models in the ensemble, $\text{GAMLSS}_{LN}$ tends to have better performance in more recent accident periods as its weight increases when data from later accident periods are added to the validation set. Indeed, as per Figure \ref{WeightsADLPSplitS2}, $\text{GAMLSS}_{LN}$ is allocated with the largest weight in the second subset. However, the weight allocated to $\text{GLM}_{\text{PPCF}}$ has demonstrated an opposite pattern: there is a substantial weight given to $\text{GLM}_{\text{PPCF}}$ with the earliest split point, whereas the weight decreases to a small value as the split point increases. This phenomenon might be explained by the ability of $\text{GLM}_{\text{PPCF}}$ to capture claims finalisation information. Since claims in earlier accident periods are close to being finalised, $\text{GLM}_{\text{PPCF}}$ tends to yield better predictive performance in earlier accident periods by taking this factor into account. This property of PPCF models has also been noted by \citet*[Chapter 5 of][pp. 151--165]{Tay00}.

Besides analysing individual models, we also analyse the weights' pattern at the group level by clustering the component models based on their structure and distributional assumptions. In terms of model structure, all models with standard cross-classified structures are assigned with small weights based in Figure \ref{WeightsADLPSplitS} and Figure \ref{WeightsADLPSplitS2}. Due to the relatively complex characteristics of the simulated data, the standard cross-classified models tend to perform poorly as they only consider the accident and development period effects. We have also found that only one of them will be allocated with a large weight for those models with similar structures. In fact, the models with the four largest weights in both Figure \ref{WeightsADLPSplitS} and Figure \ref{WeightsADLPSplitS2} all have different structures. This finding highlights the advantage of using linear pools as the weights of potential correlated models can be shrunk towards zero by imposing the non-negativity constraint on weights.

From the angle of distributional assumptions, all models with ODP distributional assumptions are assigned relatively small weights, whereas models with Log-Normal distributional assumptions are allocated larger weights. Due to the relatively large tail-heaviness of incremental claims' distribution, Log-Normal tends to provide a better distributional fit as it has a heavier tail than the ODP distribution. Additionally, we have also found that Zero-Adjusted Gamma distribution (ZAGA) is assigned with a larger weight when the split occurs earlier as shown in Figure \ref{WeightsADLPSplitS}. The weight allocation might be explained by the tendency for a greater proportion of zero incremental claims in earlier accident periods as the claims from earlier accident periods are usually closer to their maturity: these are the ones that are the most likely to have 0 claims in the very late development periods (the later accident years haven't reached that stage in the training set). Since the design of the ZAGA model allows the probability mass of zero incremental claims to be modeled as a function of development periods (i.e., the closeness of claims to their maturity), ZAGA tends to yield better performance on when only the most mature accident periods are used. Interestingly, the Zero-Adjusted Log-Normal (ZALN) model has been assigned with small weights for earlier accident periods splits, which might be explained by the fact that it provides overlapping information with the ZAGA model.

\subsection{Evaluating performance across varying data sizes}

To assess the robustness of our results against different data sizes, we conducted additional tests using a $20\times 20$ reserving triangle, which represents the aggregation of claims at semi-annual levels. To ensure consistency, we employed the same sets of simulation parameters from the SynthETIC generator to simulate the claims payments in both cases.

Using a similar partition strategy as the 40 x 40 case, the $20\times 20$ triangle is split into a training set, a validation set and a out-of-sample set to build the ensembles and evaluate models' performance. This split is illustrated in Figure \ref{fig:DataPartition20x20} in Appendix \ref{Appendix20x20Results}.

Following the same modelling framework as in Section \ref{modCombine}, the results for the $20\times 20$ triangle are obtained and listed in Appendix \ref{Appendix20x20Results}. In summary, the ADLP ensembles continue to out-perform the SLP, the equally weighted ensemble and the BMV in most simulations based on the Log Score, even when considering the smaller triangle size. Based on the statistical test results, the majority of $p$-values falls below the conventional $5\%$ level when assessing the score differential between the ADLP and the BMV (indicating statistical significance). Although the count of statistically significant outcomes experiences a slight reduction when assessing the performance difference between the ADLP and the equally weighted ensemble, it remains substantial, constituting over half of the total simulations. However, the difference between SLP and ADLP becomes less significant as the triangle size decreases. This suggests that (unsurprisingly) simpler data partition strategies may be suitable when dealing with relatively small triangle sizes.

As for the change in the composition of the ensembles, we have found (again unsurprisingly) that more parsimonious models receive larger weights on the $20\times 20$ triangle. Since there are no instances of zero incremental claims at higher aggregation levels, models utilising zero-adjusted distributions (specifically, ZALN and ZAGA) are excluded from the ensembles before fitting.

The results above suggest that, while it may be possible to apply our ensembling framework to even smaller triangles, the added value of doing so might not be as significant. In practice and nowadays, we expect that most companies would be in a position to analyse data at sufficient level of granularity in order to reap the potential benefits of the approach developed in this paper.

\section{Conclusions} \label{conclusions}

In this paper, we have designed, constructed, analysed and tested an objective, data-driven method to optimally combine different models for reserving. Using the Log Score as the combination criterion, the proposed ensemble modelling framework has demonstrated its potential in yielding high distributional forecast accuracy.

Although the standard ensembling approach (e.g. the standard linear pool) has demonstrated success in non-actuarial fields, it does not yield the optimal performance in the loss reserving context considered in this paper. To address the common general insurance loss data characteristics as listed in Section \ref{StatementContri}, we introduced the \textbf{Accident and Development period adjusted Linear Pools (ADLP)} modelling framework.   This includes a novel validation strategy that takes account into the triangular loss reserving data structure, impacts and features related to different Accident and Development period combinations.

In particular, the partitioning strategy not only effectively utilises variations in component models' performance in different accident periods, but also captures the impact from late development periods on incremental claims, which can be particularly important for long-tailed lines of insurance business. 
Based on the illustrative examples which use  simulated aggregate loss reserving data, the ADLP ensembles outperform traditional approaches (i.e., the model selection strategy BMV and the equally weighted ensemble EW) and the standard linear pool ensemble, when the Log Score, CRPS, or relative reserve bias at the $75^{th}$ quantile are used for the assessment. 

Furthermore, we also investigate the impact of a split point on the ensemble's performance. Given the composition of models included in the ensemble, we have found that splitting data into two subsets close to the middle point (i.e., between accident period 15 and 20 for a 40 x 40 loss triangle) can generally yield satisfactory performance by taking the advantages of several high-performing models in earlier accident periods while maintaining sufficient data size for training the combination weights in the both subsets.

\section*{Acknowledgements}

Authors are very grateful for comments from two anonymous referees, which led to significant improvements of the paper. 

Earlier versions of this paper were presented at the Actuaries Institute 2022 All Actuaries Summit, 
the 25th International Congress on Insurance: Mathematics and Economics (Online),
the 2022 Virtual ASTIN/AFIR Colloquium,
the One World Actuarial Research Seminars,
\rev{as well as at the 2023 CAS Annual Meeting in Los Angeles}. The authors are grateful for constructive comments received from colleagues who attended those events. 

This research was supported under Australian Research Council's 
Discovery Project funding scheme (DP200101859). The views expressed herein are those of the authors and are not necessarily those of the supporting organisations. Additionally, the authors are grateful for the research assistance of Yun Wai (William) Ho, who contributed materially to the review of the code related to the examples presented.  

\rev{The authors are also grateful to the Casualty Actuarial Society (CAS) for awarding the 2023 Hachemeister Prize to this paper, and for supporting its presentation at the 2023 CAS Annual Meeting in Los Angeles.}

\section*{Data and Code}
All data sets were generated using the CRAN package \texttt{SynthETIC} as explained above. \rev{Results presented in this paper can be replicated using the R code available at} \url{https://github.com/agi-lab/reserving-ensemble}.

\section*{\rev{R package \texttt{ADLP}}}

\rev{An R package entitled \texttt{ADLP} has been developed based on the Accident and Development period adjusted Linear Pools (ADLP) method, and is available from CRAN (link: \url{https://cran.r-project.org/web/packages/ADLP/index.html}). It features a suite of functions for calibrating ADLP ensembles, calculating evaluation metrics, and generating simulations from fitted ADLP objects. Additionally, the package grants users ample flexibility to choose or create component models for the ensemble, and to employ data partitioning for calibrating either the component models or the combination weights, aligning with their experiences.}

\section*{Declarations of interest}

None

\section*{References}

\bibliographystyle{elsarticle-harv}
\bibliography{libraries}

\appendix

\section{Descriptions of component models} \label{Appendix:ComponentModels}

\subsection{$\text{GLM}_{\text{CC}}$: the basic form} \label{CCGLM}

The first model included in our ensemble is the GLM with cross-classified structure that has been introduced in modelling outstanding claims since \citet*{ReVe98}. The standard cross-classified models can be fitted mechanically, and there is usually no manual adjustment needed for tuning the parameters. Using the notations from \citet*[Chapter 2 of][pp. 8--12]{TaMG16}, the linear predictor can be specified as
\begin{eqnarray} \label{GLMCCFormula}
\eta_{i,j} &=& \ln(\alpha_{i})+\ln(\beta_{j}),
\end{eqnarray}
where the accident period effect and development period effect can be captured by the parameters $\ln(\alpha_{i})$ and $\ln(\beta_{i})$, respectively. Using the log-link function (i.e. $\mu_{ij}=\exp(\eta_{i,j})$), the predicted incremental claim is modelled as 
\begin{eqnarray}
\hat{\mu}_{ij} &=& \hat{\alpha}_{i} \cdot \hat{\beta}_{j}.
\end{eqnarray}
Therefore, the cross-classified models have the nice interpretation that the predicted incremental claim is the product of the estimated total amount of claims paid in accident period $i$ (i.e. $\hat{\alpha}_{i}$) and the proportion of that total claims paid in development period $j$ (i.e. $\hat{\beta}_{j}$). $\text{GLM}_{\text{CC}}$ is fitted under the Over-Dispersed Poisson (ODP), Gamma, and Log-Normal distribution, which are three common distribution assumptions in loss reserving literature, respectively. The above distribution assumptions are summarised below:

\begin{itemize}
    \item ODP: $\text{E}[Y_{i,j}]=\exp(\eta_{i,j})$ and $\text{Var}(Y_{i,j})=\phi\mu_{i,j}$;
    \item Log-Normal: $\text{E}[\ln(Y_{i,j})]=\eta_{i,j}$ and $\text{Var}(\ln(Y_{i,j}))=\sigma^{2}$;
    \item Gamma: $\text{E}[Y_{i,j}]=\mu_{i,j}=\exp(\eta_{i,j})$ and $\text{Var}(Y_{i,j})=\phi \mu_{i,j}^{2}$.
\end{itemize}

Note that $\text{GLM}_{\text{CC}}$ with the ODP distribution assumption yields the identical central estimate 
of outstanding claims as to the traditional Chain-Ladder algorithm \citep*{ReVe98}.

\subsection{$\text{GLM}_{\text{Cal}}$: incorporating calendar period effect} \label{CalPeriod}

As \citet*[Chapter 7 of][pp. 76--94]{TaMG16} commented, calendar period effects can be present due to economic inflation or changes in claims management. Therefore, the second model seeks to capture this effect by introducing the parameter $\gamma$ in the linear predictor \citep*{TaXu16}: 
\begin{eqnarray} \label{GLMCalFormula}
\eta_{i,j} &=& \ln(\beta_{j})+t \cdot \ln(\gamma). 
\end{eqnarray}
The calendar period effect is assumed to be constant such that the model can be applied mechanically without much users' adjustment based on Criterion 1. $\text{GLM}_{\text{Cal}}$ is fitted using the same distribution assumptions as $\text{GLM}_{\text{CC}}$. 

The model with calendar period effect has a close connection with the age-period-cohort model proposed by \cite*{KuNiNi08} and \cite*{HaNi18}. If we assume a separate effect for each calendar period and add the accident period effects, the model defined in \ref{GLMCalFormula} will become:
\begin{eqnarray} \label{GLMCalFormula2}
\eta_{i,j} &=& \ln(\alpha_{i})+\ln(\beta_{j})+ \ln(\gamma_t),  
\end{eqnarray}
which has the same structural form as the age-period-cohort type of models in \cite*{KuNiNi08} and \cite*{HaNi18}. Furthermore, removing the period term in \eqref{GLMCalFormula2} transforms the model into an age-cohort model \citep{MiDoBePe15}, which also corresponds to the cross-classified model defined in \eqref{GLMCCFormula} (i.e., $\text{GLM}_{\text{CC}}$). However, it is also important to recognise that the forecasting methodologies for the models specified in \eqref{GLMCalFormula} and \eqref{GLMCalFormula2} differ. Specifically, for forecasting purposes, the period parameter (denoted as $\ln(\gamma_t)$) in \eqref{GLMCalFormula2} requires extrapolation to produce forecasts for the lower triangle. In contrast, the $\text{GLM}{\text{Cal}}$ model, as defined in \eqref{GLMCalFormula}, treats the calendar parameter as a numeric variable, which does not require this additional extrapolation step.

\subsection{$\text{GLM}_{\text{HC}}$: incorporating the Hoerl Curve} \label{HoerlCurve}

The third GLM seeks to capture the shape of development pattern of incremental claims by incorporating a Hoerl curve \citep*{Wri90} in the linear predictor, which can be specified as
\begin{eqnarray}
\eta_{i,j} &=& a_{i}+b \ln(j)+cj.
\end{eqnarray}
The number of parameters in $\text{GLM}_{\text{HC}}$ for a 40x40 loss triangle have been significantly reduced from 80 to 42 compared with $\text{GLM}_{\text{CC}}$. By adding the Hoerl curve to the ensemble, the potential over-fitting risk of the cross-classified GLM can be mitigated. Similarly, $\text{GLM}_{\text{HC}}$ is fitted under the ODP, Gamma, and Log-Normal distribution assumptions. 

\subsection{$\text{GLM}_{\text{PPCI}}$: incorporating reported claims count information} \label{PPCIMethod}

Incorporating reported (or notified) claims count might lead to better data representation as claim payments could be subject to the impact of claims notification \citep*[Chapter 7 of][pp. 76--94]{TaMG16}, particularly for lines of business with volatile payments but stable average payments per claim \citep*{TaXu16}. Following the notations from \citet*{TaXu16}, the linear predictor can be specified as 
\begin{eqnarray}
\eta_{i,j} &=& \ln(\hat{N}_i)+\ln(\beta_{j}), 
\end{eqnarray}
where $\hat{N}_i$ denote the estimated total reported claims count for accident period $i$. $\hat{N}_i$ can be specified as $\hat{N}_{i}=\sum_{j=1}^{J-i+1}N_{ij}+\sum_{j=J-i+2}^{J}\hat{N}_{ij}$, where $\hat{N}_{ij}$ denote the reported claims count that can not be observed and therefore need to be estimated. Following the methodology in \citet*{TaXu16}, $\hat{N}_{ij}$ is estimated from the reported claims count triangle by $\text{GLM}_{\text{CC}}$ with Over-Dispersed Poisson distribution assumption. $\text{GLM}_{\text{PPCI}}$ is fitted under the ODP distribution assumption.

\subsection{$\text{GLM}_{\text{PPCF}}$: incorporating finalised claims count information} \label{PPCFMethod}

Incremental claims can also be impacted by the claims closure rate, which can be captured by the PPCF model \citep*{TaXu16}. Following the methodology from \citet*{TaXu16}, the payment per finalised claim, defined as $Y_{ij}/\hat{F}_{ij}$, is modeled by an Over-dispersed Poisson GLM:
\begin{eqnarray}
\hat{\mu}_{ij}^{\text{PPCF}} &=& \exp(\beta_0+\beta_1\hat{t}_i(j)),
\end{eqnarray}
where $\hat{t}_i(j)$ is the (estimated) proportion of claims closed up to development period $j$ that are incurred in accident period $i$, and where $\hat{F}_{ij}$ is estimated by a binomial GLM \citep*{TaXu16}. Finally, the estimated incremental claim can be derived as: $\hat{Y}_{ij}=\hat{\mu}_{ij}^{\text{PPCF}} \cdot \hat{F_{ij}}$

\subsection{$\text{GLM}_{\text{ZALN}}$ and $\text{GLM}_{\text{ZAGA}}$: Handling Incremental Claims with Zero Values} \label{HandleZeros}

In addition to the Over-Dispersed Poisson, Log-Normal, and Gamma distribution assumptions, $\text{GLM}_{\text{ZALN}}$ and $\text{GLM}_{\text{ZAGA}}$ are fitted using the Zero-Adjusted Log-Normal (ZALN) distribution and the Zero-Adjusted Gamma (ZAGA) distribution, which are specialised at handling zero incremental claims by assigning a point mass at zero \citep*{ReIsJa13}. Denote the probability mass at zero as $\nu_j$. We assume the point mass to be dependent on development periods, which takes into account the fact that zero incremental claims are more likely to be observed in later development periods when the claims are close to being settled.
$\nu_j$ can be then modeled by fitting a logistic regression: 
\begin{equation}
\hat{\nu}_j=\frac{\exp(\beta_0+\beta_1j)}{1+\exp(\beta_0+\beta_1j)}.
\end{equation}
Denote $\hat{\mu}_{ij}^{\text{GA}}$ and $\hat{\mu}_{ij}^{\text{LN}}$ as the predicted mean for Gamma and Log-Normal distribution, respectively. Finally, the predicted incremental claims under the ZAGA and ZALN distribution can be specified as $\hat{\mu}_{ij}^{\text{ZAGA}}=(1-\hat{\nu}_j)\hat{\mu}_{ij}^{\text{GA}}$ and $\hat{\mu}_{ij}^{\text{ZALN}}=(1-\hat{\nu}_j)\hat{\mu}_{ij}^{\text{LN}}$, respectively. 

\subsection{Smoothing Spline} \label{SP}

Smoothing splines provide a flexible approach to automatically balance data adherence and smoothness. Comparing to fixed parametric curves, smoothing splines have a more flexible shape to incorporate different patterns present in claims development \citep*{EnVe01}. Following the work in \citet*{EnVe01}, the linear predictor can be specified as
\begin{eqnarray} \label{SPmean}
\eta_{ij} &=& s_{\theta_{i}}(i)+s_{\theta_{j}}(j),
\end{eqnarray}
where $s_{\theta_{i}}$ are $s_{\theta_{j}}$ are smoothing functions over the accident period $i$ and development period $j$. The distributional assumption of the smoothing splines implemented in this paper are Normal, Gamma and Log-Normal. The Normal distribution assumption is the original distributional assumption for smoothing spline in \citep*{EnVe01}, where the latter two distribution assumptions are commonly used in loss reserving literature. For each distributional assumption, the smoothing functions for the accident periods and the development periods are found by maximizing their respective penalised log-likelihoods \citep*{GrSi93}:
\begin{eqnarray}
l(s_{\theta_i}) &=& \sum_{i,j} \left(\frac{y_{ij} s_{\theta_i}(i)-\exp(s_{\theta_i}(i))}{\phi}\right)-\theta_{i}\int \left(s_{\theta_i}''(t)\right)^2 dt \\
l(s_{\theta_j}) &=& \sum_{i,j} \left(\frac{y_{ij} s_{\theta_j}(j)-\exp(s_{\theta_j}(j))}{\phi}\right)-\theta_{j}\int \left(s_{\theta_j}''(t)\right)^2 dt;
\end{eqnarray}
where $\theta_{i}$ and $\theta_{j}$ are hyper-parameters that control the degree of smoothness for smoothing spline. The parameters $\theta_{i}$ and $\theta_{j}$ are estimated by minimizing the Generalised Approximate CV Score (GACV) \citep*{GuXi01}.

\subsection{GAMLSS: Generalised Additive Models for Location Scale and Shape} \label{GAMLSS}

Under the standard GLM setting, the dispersion parameter is assumed to be constant across all accident periods and development periods. However, dispersion levels typically change in practice across development (and sometimes accident) periods \citep*[Chapter 6 of][pp. 56--72]{TaMG16}. By modelling the variance as a function of predictors, GAMLSS, which is firstly proposed by \citet*{RiSt05}, might tackle the inconstant dispersion issue automatically \citep*{Sp14}. Therefore, it is included in our ensembles.

Following the work from \citet*{Sp14}, the linear predictor for the distribution mean is in the standard cross-classified form. Although the distribution variance can also be specified in the cross-classified form, to mitigate the potential over-parametrisation risk, a smoothing spline is applied to the development period when modelling the variance term. We also assume the distribution variance to vary only by development periods, which matches the common assumption in loss reserving literature \citep*[Chapter 6 of][pp. 56--72]{TaMG16}. Under the above assumptions, the mean and the variance of the distribution of incremental claims can be modelled as
\begin{eqnarray} 
\text{E}[Y_{ij}] = g^{-1}(\eta_{1,i,j})  &=& \exp(\ln(\alpha_i)+\ln(\beta_j)); \label{GAMLSSMean} \\
\text{Var}[Y_{ij}] = g^{-1}(\eta_{2,i,j}) &=&  \exp(s_{\theta_{j}}(j)); \label{GAMLSSDisper}
\end{eqnarray}
where $s_{\theta_{j}}$ is a smooth function applied here to the development period $j$ only. In this paper, the GAMLSS model is fitted under the distribution assumption of Gamma and Log-Normal, which are common distribution assumptions for loss reserving models.

\section{Distributional forecasting concepts}

\subsection{Log Score and Kullback-Leibler Divergence Distance} \label{ALIC}

The Log Score has a close connection to the Kullback-Leibler divergence (KLIC) distance between the true density $f_m$ and the predicted density $\hat{f}_m$ \citep*{HaMi07}, which is defined as: 
\begin{eqnarray}
\text{KLIC} &=& \ln(f_m(y_{ij}))-\ln(\hat{f}_{m}(y_{ij})) 
\end{eqnarray}
The smaller the KLIC distance, the smaller the deviation of the predictive density from the true density. If the predictive density is equal to the true density, its KLIC distance will be zero. However, direct minimization of the difference between the true density and the predictive density might not be feasible, as the true density is often unknown in practice. One solution is to maximise $\ln(\hat{f}_{m}(y_{ij}))$, which is equivalent to minimise the KLIC distance \citep*{HaMi07}. As $\ln(\hat{f}_{m}(y_{ij}))$ corresponds to the Log Score of the predictive density, the logarithmic scoring rule has the nice interpretation of assigning a higher score to the predictive density with a closer distance to the true density \citep*{HaMi07}. 

\section{Technical details}
 
\subsection{Details of Calculating Reserve Quantiles} \label{CalReserQuant}

To simulate the $75^{th}$ reserve quantile for component model $m$, we implement the following procedure: 
\begin{itemize}
    \item 1. Simulate $N$ random variables for each cell in out-of-sample data from its corresponding distribution; Denote $\tilde{Y}_{ij}^{m}=(\tilde{Y}_{ij,(1)}^{m},\tilde{Y}_{ij,(2)}^{m},...,\tilde{Y}_{ij,(N)}^{m})$ as the vector of all simulated random variables for cell (i,j) under the component model $m$
    \item 2. Repeat Step 1 for each cell $(i,j)$
    \item 3. Calculate the simulated reserve based on the simulated random variables for each cell $(i,j)$: $\tilde{R}^m=(\tilde{R}^m_{(1)},\tilde{R}^m_{(2)},...,\tilde{R}^m_{(N)})=(\sum_{i,j\in D^{out}}\tilde{Y}_{ij,(1)}^{m},\sum_{i,j\in D^{out}}\tilde{Y}_{ij,(2)}^{m},...,\sum_{i,j\in D^{out}}\tilde{Y}_{ij,(N)}^{m})$
    \item 4. Calculate the empirical $75^{th}$ quantile of the $N$ simulated reserves for component model $m$: $R^{m}_{75}$
\end{itemize}
Here we take $N=1000$, which provides enough to be computationally feasible but also reasonably accurate at the quantiles we are looking at. One should also note that the above simulation procedure implicitly assumes the independence of individual cells, which is a common assumption used to simulate reserve quantiles in literature \citep*{Gab20}. The procedure mentioned above is similar to the parametric bootstrapping method of simulating reserve \citep*[Chapter 5 of][pp. 42--53]{TaMG16}. Since this paper focuses on assessing predictive distributions, we do not pursue non-parametric or semi-parametric bootstrapping further here. 

\subsection{Constraints on Combination Weights: Acting as Downside Insurance} \label{proofConstraints}

As \citet{Br96} suggests, imposing the sum-to-unity and non-negativity constraints act as downside insurance for the out-of-sample predictive performance of the ensemble. This proposition can be supported by the following simple proof.

Denote $\hat{f}^{*}(y_{ij})=\sum_{m=1}^{M}w_m\hat{f}_m(y_{ij})$ as the predictive density for the ensemble at out-of-sample point $y_{ij}$, and $\hat{f}_{worst}(y_{ij})=\text{min} \{\hat{f}_m(y_{ij}),m=1,...,M\}$ as the the lowest predictive density at the same data point from one of the component model m (i.e. the predictive density from the worst component model). By imposing the sum-to-unity constraint and non-negativity constraints on model weights, $\ln(\sum_{m=1}^{M}w_m\hat{f}_m(y_{ij})) \ge \ln(\hat{f}_{worst}(y_{ij}))$ is guaranteed.

\subsection{Comparison between Stacked regression approach and linear pools: further illustrations} \label{Mean-var-tradeoff}

Section \ref{StackedregressionDiscuss} states that the MSE does not give forecasters the incentive to model higher moments correctly. We supplement the discussion in Section \ref{StackedregressionDiscuss} with the following illustrative example. Suppose the true distribution for cell $(i,j)$ is $\text{N}(2,2)$. Consider the predictive distribution $\text{N}(2,1)$. Since it has the same mean as the true distribution, it will receive the same MSE as the true distribution, even though it severely underestimates the true dispersion. Consequently, the estimated $75^{\text{th}}$ quantile from the forecasting distribution, calculated as $\hat{\mu}+\hat{\sigma} z_{0.75}$, will also be lower than the true $75^{\text{th}}$ quantile. The potential of under-estimation of higher quantiles could be of particular concern to general insurers and regulators. 

As mentioned in Section \ref{StackedregressionDiscuss}, optimising the model's performance on central estimation might harm the variance forecast. To illustrate this idea, we follow the decomposition of Mean-Squared-Error and prediction variance given in \cite*{KnKr22}. Firstly, the weighted average Mean-Squared Error attained by component models can be decomposed as: 
\begin{equation} \label{WAMSE}
\begin{split}
    \sum_{m=1}^{M} w_m \text{E}[(Y-\mu_m)^2] & = \sum_{m=1}^M w_m \text{E}[(Y-\mu^{*}+\mu^{*}-\mu_m)^2] \\
    & = (\sum_{m=1}^M w_m) \text{E}[(Y-\mu^{*})^2] + \sum_{m=1}^M w_m \text{E}[(\mu_m-\mu^{*})^2] \\
    & = \text{E}[(Y-\mu^{*})^2] + D,
\end{split}
\end{equation}
where $D = \sum_{m=1}^M w_m E[(\mu_m-\mu^{*})^2]$ is the disagreement term, and $\sum_{m=1}^M w_m = 1$. Assume the individual variance forecast is unbiased (i.e., $\text{E}[\sigma_m^2]=\text{E}[(Y-\mu_m)^2]$). Then the Mean-Squared Error attained by the ensemble can be specified as:
\begin{equation} \label{EnsembleMSE}
\begin{split}
\text{E}[(Y-\mu^{*})^2] &= \sum_{m=1}^{M} w_m \text{E}[(Y-\mu_m)^2]-D  = \text{E}[\sigma_m^2]-D,
\end{split}
\end{equation}
and the (expected value of) predicted variance under the combined distributional forecast is:
\begin{equation} \label{EnsembleVAR}
\begin{split}
\text{E}[\tilde{\sigma}^2] & = \sum_{m=1}^M  w_m  \text{E}[\sigma^2_m] + \sum_{m=1}^M  w_m \text{E}[(\mu_m - \mu^{*})^2]  = \sum_{m=1}^M  w_m  \text{E}[\sigma^2_m] + D.
\end{split}
\end{equation}

Equation \eqref{EnsembleMSE} and \eqref{EnsembleVAR} have several important implications. To obtain the optimal central estimate, one should minimise the Mean-Squared error attained by the ensemble. As per \eqref{EnsembleMSE}, the MSE of the combined forecast decreases as the disagreement term increases. Therefore, if our goal is to achieve optimal central estimates, the optimal weights should be derived such that the disagreement among component models could be maximised, which is usually the case under the stacked regression approach. However, as per \eqref{EnsembleVAR}, the predicted variance increases with the disagreement term. Therefore, if the individual forecast variances are unbiased or over-estimated, the optimal weights derived by minimising the MSE will result in a further over-statement of the ensemble's predicted variance \citep*{KnKr22}.

\section{Additional fitting results}

\subsection{ADLP with Three Subsets} \label{ADLPThreeSubsets}

The average out-of-sample Log Score attained by the four $\text{ADLP}^{+}$ ensembles, in comparison with the corresponding ADLP ensembles, are summarised in Table \ref{LSCompTwovsThree}. The ``Split Points" column shows the accident period(s) where the split of subsets occurs for both ADLP ensembles (outside the bracket) and $\text{ADLP}^{+}$ ensembles (inside the bracket). As per Table \ref{LSCompTwovsThree}, the average Log Score differential between the ADLP and the $\text{ADLP}^{+}$ ensembles tend to be small, except for $\text{ADLP}_9$ and $\text{ADLP}_9^{+}$. 

\begin{figure}[htb]
    \centering
    \includegraphics[width=0.5\textwidth]{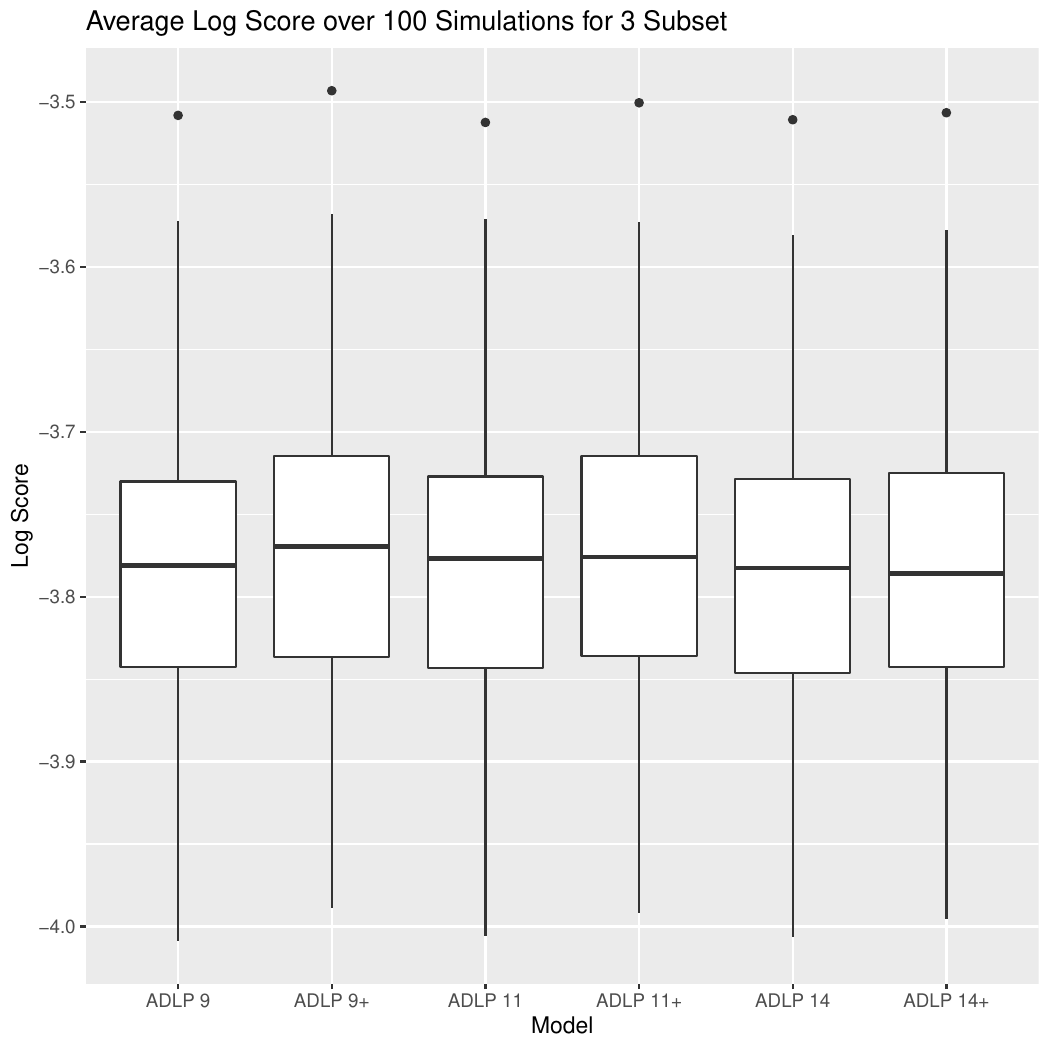}
    \caption{Box-plot of Log Score: ADLP v.s. $\text{ADLP}^{+}$}
    \label{BoxplotTwovsThree}
\end{figure}

Additionally, based on the box-plot of Log Score in Figure \ref{BoxplotTwovsThree}, the distributions of Log Score for the $\text{ADLP}$ and $\text{ADLP}^{+}$ over the 100 simulations are also similar, though $\text{ADLP}_9$ under-performs $\text{ADLP}_9^{+}$ in most simulated data set by a small margin.

\begin{table}[htb]
    \centering
    \caption{Comparison of Average Log Score: ADLP v.s. $\text{ADLP}^{+}$ \label{LSCompTwovsThree}}
    \begin{tabular}{llll}
Split Points             & Average Log Score (ADLP)     & Average Log Score ($\text{ADLP}^{+}$)  \\ \hline
AP15(AP15 and 29)   & 	-3.7844          &   -3.7731         \\
AP17(AP17 and 31) &  -3.7837        &   -3.7755 \\ 
AP23(AP23 and 33) &  -3.7871	        &  -3.7843          \\ \hline
    \end{tabular}
\end{table}

As per Table \ref{LSCompTwovsThree}, $\text{ADLP}^{+}_9$ has the highest Log Score among all ADLP ensembles with three subsets, and $\text{ADLP}_{11}$ yields the best performance among the two-subsets ADLP ensembles. We also test whether the Log Score differential between those two partition strategies is statistically significant. Under the $5\%$ significance level, the rejection rate is considered low. Since $\text{ADLP}^{+}_9$ is the best $\text{ADLP}^{+}$ ensembles based on Log Score, the test results also imply any other three-subsets partition strategy cannot statistically significantly improve $\text{ADLP}_{11}$. Additionally, we also feel parsimony would suggest restricting ourselves to a single split point.

\subsection{Fitting results for a reserving triangle of dimension $20\times 20$} \label{Appendix20x20Results}

\subsubsection{Data partition under the $20\times 20$ triangle case}

\begin{figure}[H]
    \centering
    \includegraphics[width=0.7\textwidth]{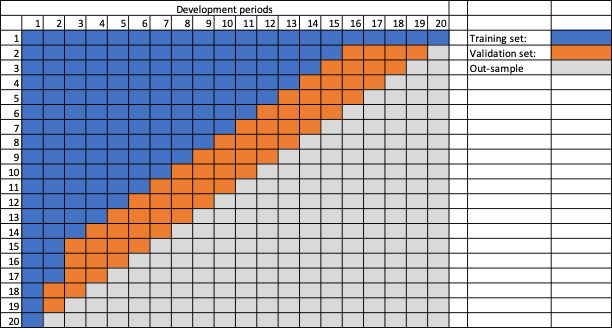}
    \caption{Data partition under the $20\times 20$ triangle case}
    \label{fig:DataPartition20x20}
\end{figure}

\subsubsection{Assessing predictive performance: Log Score}

To evaluate the impact of partition strategies on the predictive performance of ADLP ensembles within a $20\times 20$ reserving triangle, we present the mean Log Score achieved for various split points below:

\begin{figure}[H]
    \centering
    \includegraphics[width=0.5\textwidth]{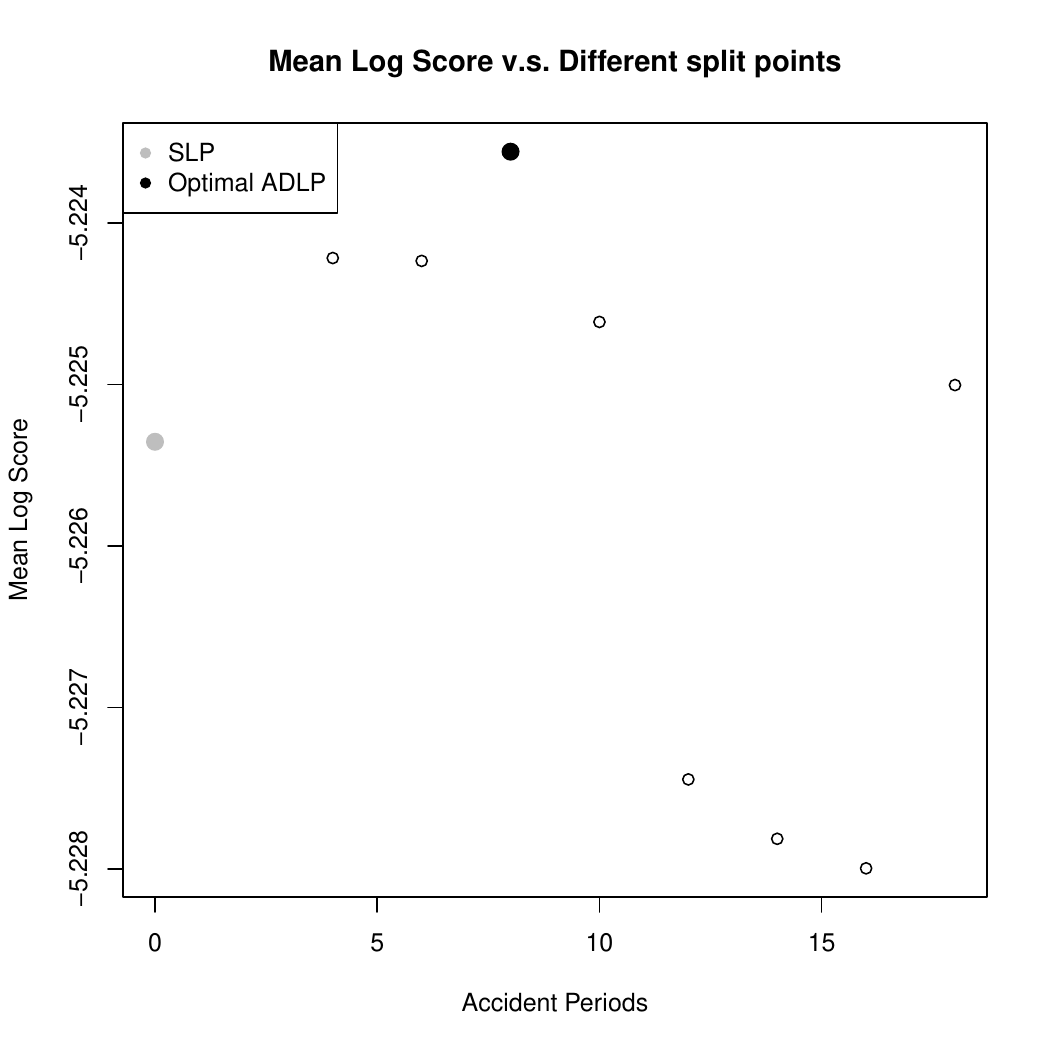}
    \caption{Mean Log Score Plot: comparison among different partition strategies ($20\times 20$ triangle)}
    \label{fig:MeanLSPartition20x20}
\end{figure}

The predictive performance of the optimal ADLP identified in Figure \ref{fig:MeanLSPartition20x20} is then compared against the SLP and the benchmark strategies: 

\begin{figure}[H]
\begin{minipage}{\textwidth}
\begin{minipage}[b]{0.5\textwidth}
    \centering
    \includegraphics[width=\textwidth]{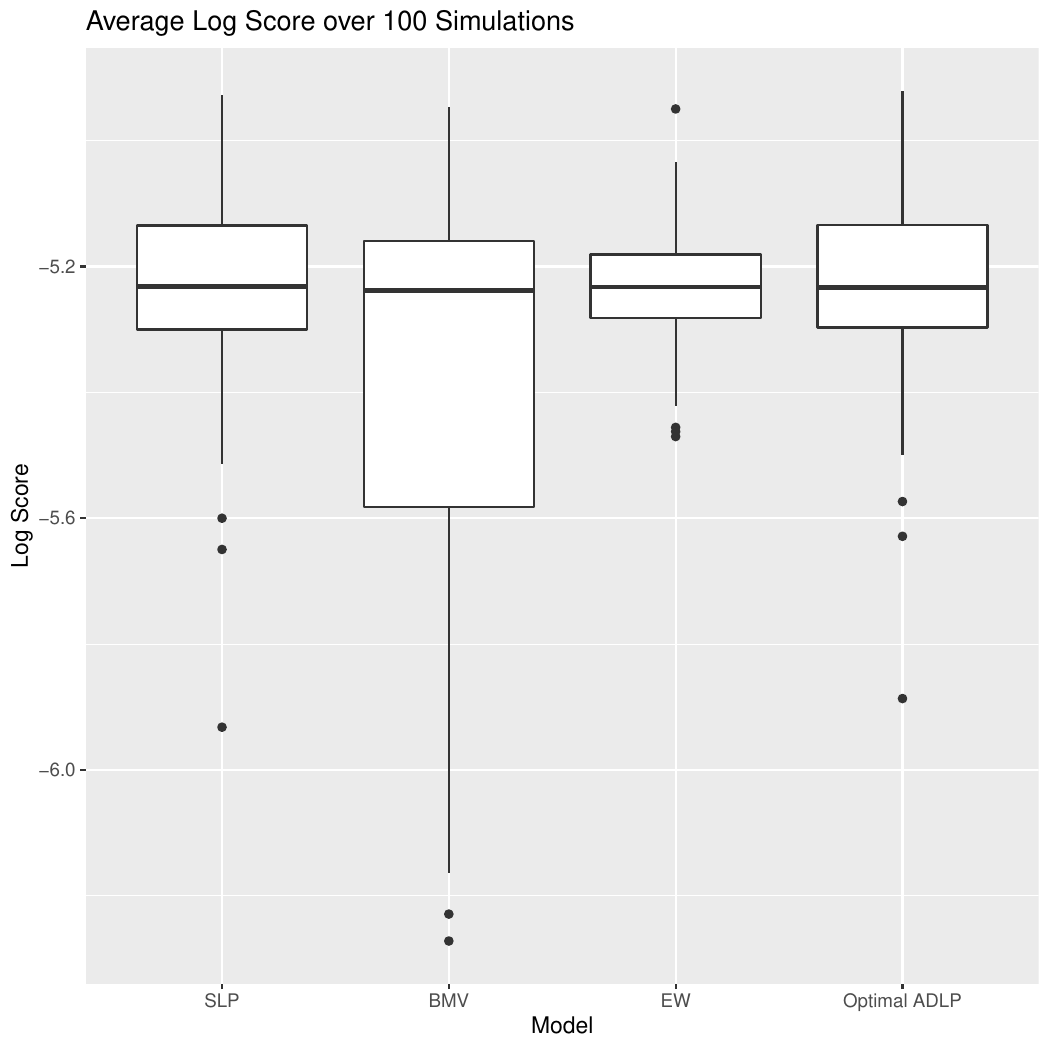}
    \captionof{figure}{Distribution of Log Score over 100 simulations ($20\times 20$ triangle): comparison among ADLP, SLP, BMV and EW}
    \label{DistLS20x20}
    \end{minipage}
  \begin{minipage}[b]{0.5\textwidth}
    \centering
    \includegraphics[width=\textwidth]{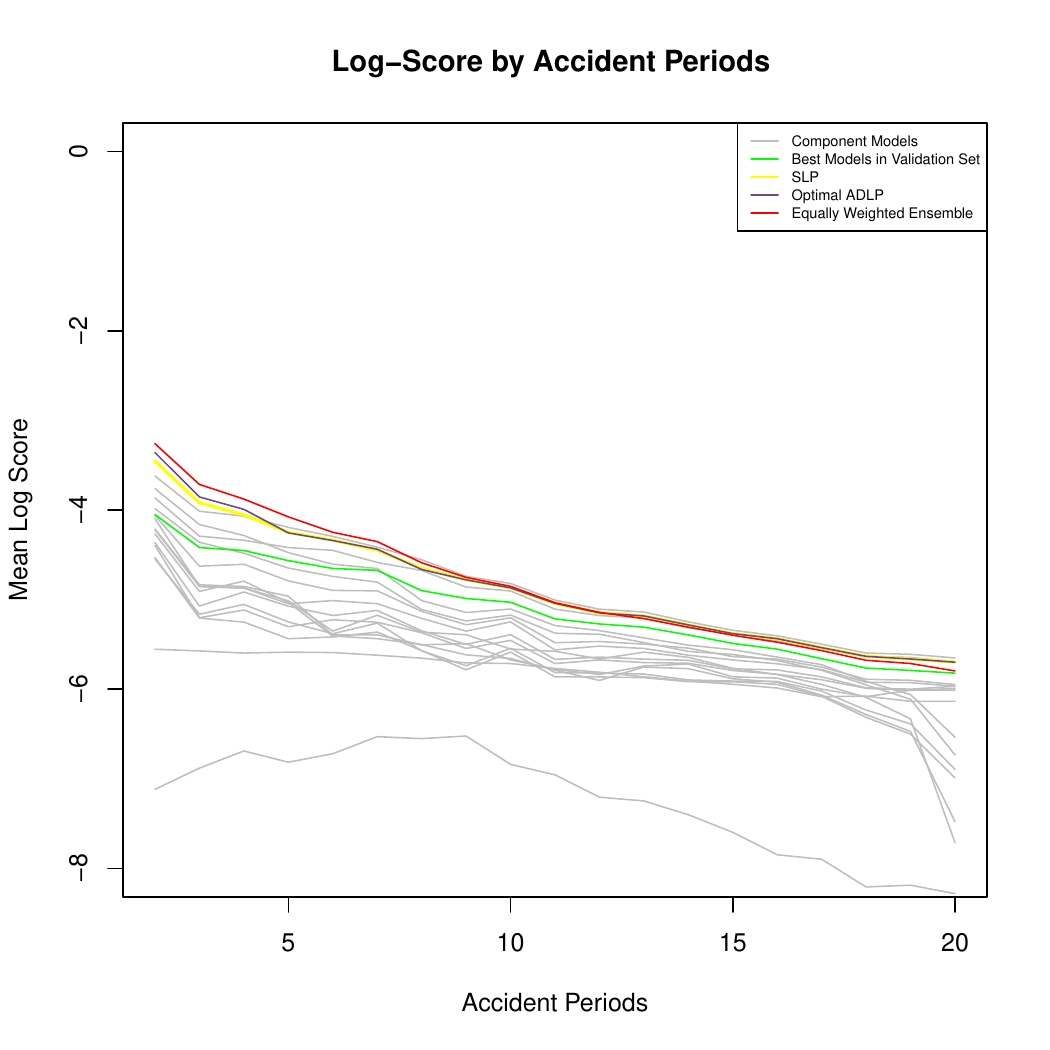}
    \captionof{figure}{Mean Log Score by accident periods ($20\times 20$ triangle)}
    \label{MeanLS20x20} 
  \end{minipage}
  \hfill
  \end{minipage}
\end{figure}

For better presentation, the mean Log Scores for the four competing strategies are summarised below:

\begin{table}[htb]
\centering
\caption{Comparison of Log Score: $\text{BMV}$, $\text{EW}$, $\text{SLP}$, and $\text{Optimal ADLP}$ \label{MeanLogScore20x20Triangles}}
\begin{tabular}{ccc}\hline
  Models            & Mean Log Score \\ \hline
$\text{BMV}$           & -5.3832    \\
$\text{EW}$                 & -5.2349	      \\
$\text{SLP}$                 & -5.2254     \\ 
$\text{Optimal ADLP}$                 & -5.2236    \\  \hline
\end{tabular}
\end{table}

\subsubsection{Statistical test results}

To assess the difference in the predictive performance among various competing strategies, we performed the Diebold-Mariano test, with the distribution of p-values shown below:

\begin{figure}[H]
    \centering
    \includegraphics[width=0.5\textwidth]{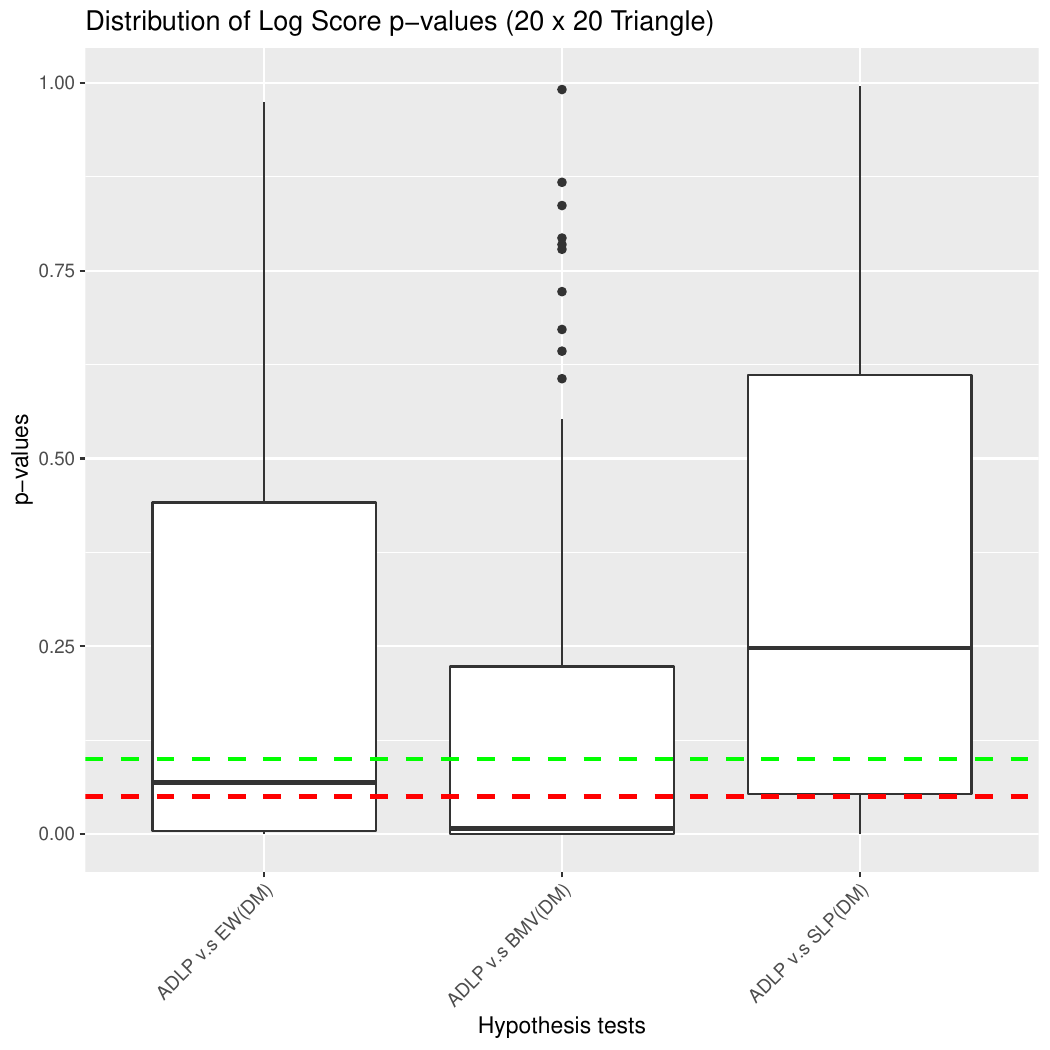}
    \caption{Distribution of p-values: ADLP v.s BMV, ADLP v.s EW, and ADLP v.s SLP; Green: marks the $10\%$ level; Red: marks the $5\%$ level}
    \label{fig:TestStatistic20x20}
\end{figure}

\begin{table}[htb]
\centering
\caption{The Number of Datasets (out of 100) when $H_0$ is rejected under the $5\%$ and $10\%$ significance levels (20 x 20 Triangle), based on the Diebold-Mariano test \label{NumDatasetsSig20x20Tri}}
\begin{tabular}{llll}
 Hypothesis Tests                          & $5\%$ level     & $10\%$ level  \\ \hline
$\text{ADLP}_{12}$ vs EW             & 47          & 54               \\
$\text{ADLP}_{12}$ vs BMV            & 65          & 67  \\
$\text{ADLP}_{12}$ vs SLP             & 22          & 33               \\  \hline
\end{tabular}
\end{table}

\subsubsection{Combination weights}

\begin{figure}[H]
\begin{minipage}{\textwidth}
\begin{minipage}[b]{0.5\textwidth}
    \centering
    \includegraphics[width=\textwidth]{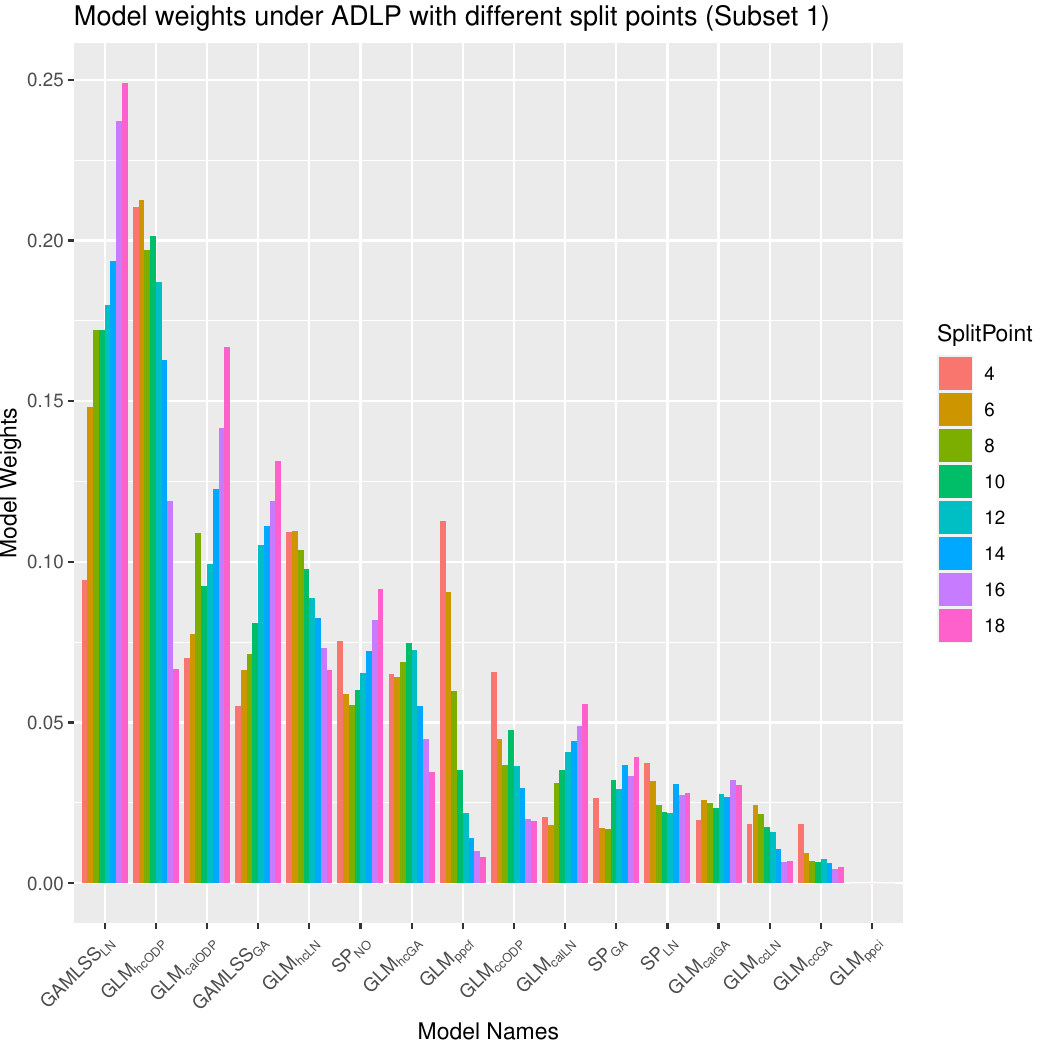}
    \captionof{figure}{Combination Weights in Subset 1 under optimal ADLP ($20\times 20$ triangle)}
    \label{CombinWeightsSubset1_20x20}
    \end{minipage}
  \begin{minipage}[b]{0.5\textwidth}
    \centering
    \includegraphics[width=\textwidth]{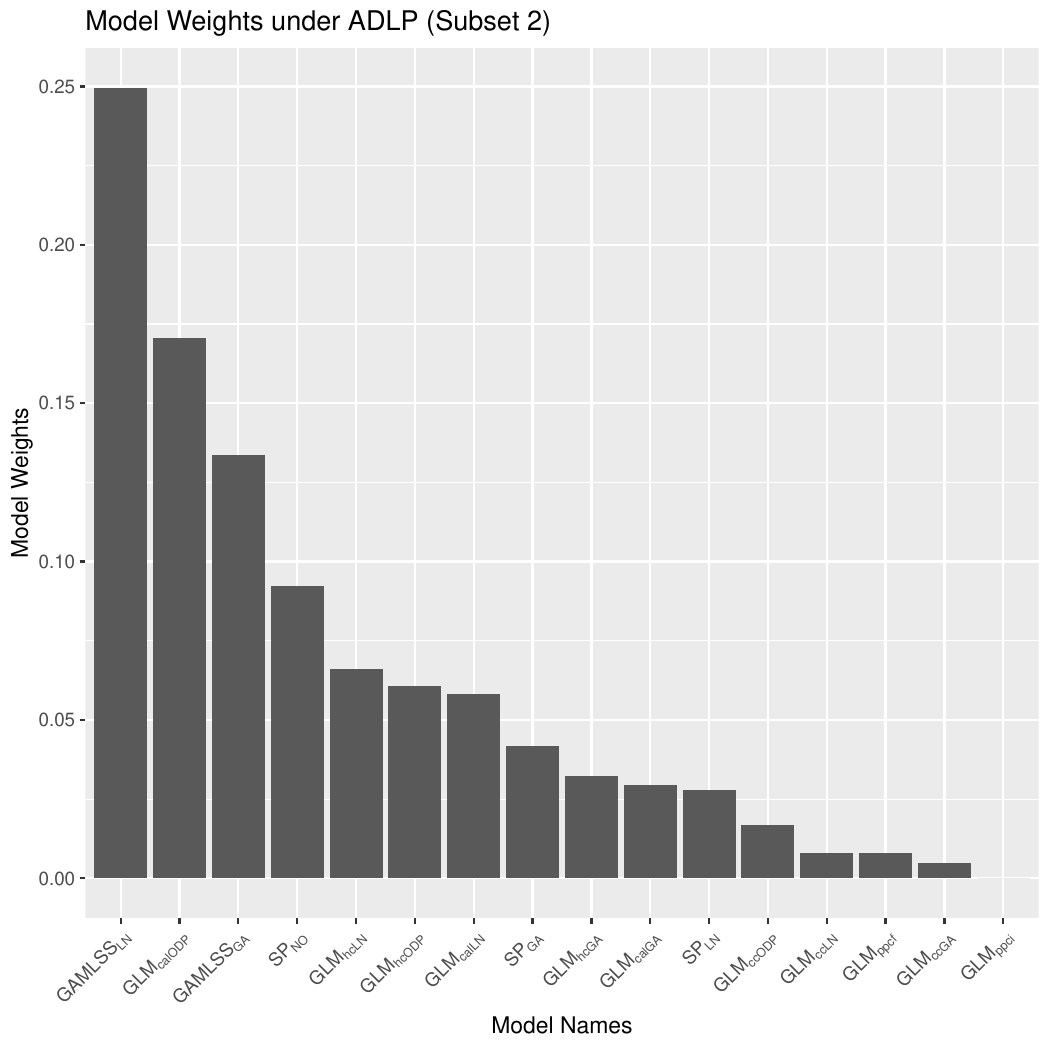}
    \captionof{figure}{Combination Weights in Subset 2 under optimal ADLP ($20\times 20$ triangle)}
    \label{CombinWeightsSubset2_20x20} 
  \end{minipage}
  \hfill
  \end{minipage}
\end{figure}

\rev{\subsection{Results for CRPS calculation} \label{Appendix:weightedCRPS}}

\rev{The CRPS can be interpreted as a measurement of the difference between the predicted and observed cumulative distributions \citep*{Her00}. Formally, the generalised version of CRPS for a predictive distribution with a cumulative distribution function $F^*$, observed at $y_k$, is defined as \citep{GnRa11b}:}
\begin{equation} \label{CRPSIntro}
    \rev{\text{CRPS}(F^*, y_k) = \int_{-\infty}^{\infty} (F^*(z)-\mathbb{I}_{z \ge y_k})^2 u(z) dz,}
\end{equation}
\rev{where $u(z)$ denotes a weight function that allows users to adjust focus on different regions of the distribution. For implementation, we use the discretised version of \eqref{CRPSIntro} \citep{GnRa11b}:
\begin{equation} \label{WeightedCRPS}
    \text{CRPS}(F^*, y_k) = \frac{z_u-z_l}{I-1} \sum_{z = z_l}^{z_u} (F^*(z)-\mathbb{I}_{z \ge y_k})^2 u(z),
\end{equation}
where $I$ is the number of discretised points, $z_l$ and $z_u$ are the lower and upper bounds for integration, respectively.}

\rev{The standard version of CRPS corresponds to the special case of $u(z)=1$. Particularly relevant to the reserving context, there is often an interest in models' performance concerning the right tail of the distribution. Thus, a weight function emphasising the right tail can be selected. To serve this purpose, a recommended choice of weight function in \citep{GnRa11b} is: $u(z) = \Phi_{a,b}(z)$, which is the cumulative distribution function of a Normal distribution with mean $a$ and standard deviation $b$. Given that $\Phi_{a,b}$ is monotonically increasing in $z$, deviations from the ideal distribution function (i.e., $\mathbb{I}_{z \ge y_k}$) are more heavily penalised as $z$ increases towards the upper end of the distribution. Inspired by the choice of the mean and standard deviation parameters for the weight function in \cite{GnRa11b}, we select $a$ to represent the average of claim payments within the lower triangle, and $b$ as the standard deviation of the claim payments in the same region. However, this choice is only one among several possibilities. In practice, actuaries might choose different weight parameters to adjust the emphasis placed on the right-tail of the distribution, guided by their expert knowledge or professional judgement.}

\begin{figure}[H]
    \centering
    \includegraphics[width=0.5\textwidth]{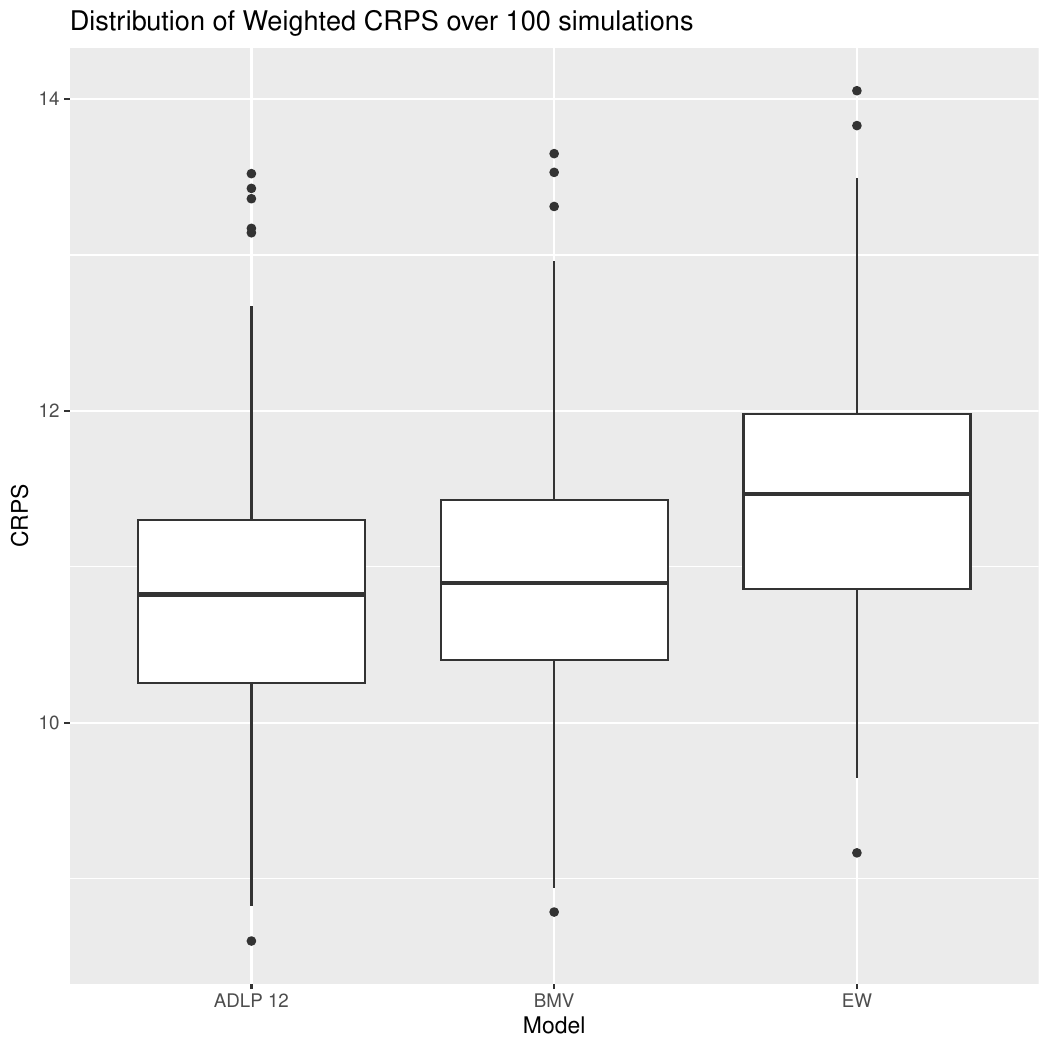}
    \captionof{figure}{Distribution of weighted CRPS (right--tail focus) over 100 simulations (lower is better): comparison among $\text{ADLP}_{12}$, EW and BMV}
    \label{BoxplotWeightedCRPS}
\end{figure}

\rev{The outcomes of the weighted CRPS calculations are presented in Figure \ref{BoxplotWeightedCRPS}. Here, $\text{ADLP}_{12}$ exhibits a noticeably lower average CRPS compared to both BMV and EW. These results, alongside the lower reserve bias observed at the $75^{th}$ quantile, suggest that ADLP's superiority over BMV and EW is more substantial in the right-tailed region of the distribution.}

\end{document}